\documentclass[aps,prc,showpacs,twocolumn,superscriptaddress,groupedaddress]{revtex4}  % for review and submission

\usepackage{graphicx}  % needed for figures\
\usepackage{dcolumn}   % needed for some tables
\usepackage{bm}        % for math
\usepackage{amssymb}   % for math
\usepackage{longtable, upgreek, amsmath}
\usepackage{hyperref}
\usepackage[caption=false]{subfig}
\begin{document}

\title{Detailed spectroscopy of $^{46}$Ca: A study of the $\beta^-$ decay of $^{46}$K }
\author{J.L. Pore} 
\altaffiliation[Present address: ]{Lawrence Berkeley National Laboratory, 1 Cyclotron Road, Berkeley, CA 94720, USA}
\affiliation{Department of Chemistry, Simon Fraser University, 8888 University Drive, Burnaby BC, V5A 1S6, Canada}

\author{C.~Andreoiu}
\affiliation{Department of Chemistry, Simon Fraser University, 8888 University Drive, Burnaby BC, V5A 1S6, Canada}

\author{J.K.~Smith}
\altaffiliation[Present address: ]{Department of Natural Sciences, Pierce College, 1601 39th Avenue SE, Puyallup, Washington 98374, USA}
\affiliation{TRIUMF, 4004 Wesbrook Mall,
Vancouver, BC, V6T 2A3, Canada}

\author{A.D.~MacLean}
\affiliation{Department of Physics, University of Guelph, 50 Stone Road E, Guelph, ON,
N1G 2W1, Canada}

\author{A.~Chester}
\altaffiliation[Present address: ]{NSCL, Michigan State University, 640 S. Shaw Lane, East Lansing, Michigan 48824, USA}
\affiliation{TRIUMF, 4004 Wesbrook Mall, Vancouver, BC, V6T 2A3, Canada}

\author{J.D.~Holt}
\affiliation{TRIUMF, 4004 Wesbrook Mall, Vancouver, BC, V6T 2A3, Canada}

\author{G.C.~Ball}
\affiliation{TRIUMF, 4004 Wesbrook Mall, Vancouver, BC, V6T 2A3, Canada}

\author{P.C.~Bender}
\altaffiliation[Present address: ]{Department of Physics, University of Massachusetts Lowell, One University Avenue, Lowell, MA 01854, USA}
\affiliation{TRIUMF, 4004 Wesbrook Mall, Vancouver, BC, V6T 2A3, Canada}

\author{V.~Bildstein}
\affiliation{Department of Physics, University of Guelph, 50 Stone Road E, Guelph, ON,
N1G 2W1, Canada}

\author{R.~Braid}
\affiliation{Department of Physics, Colorado School of Mines, 1500 Illinois Street, Golden, CO
80401, USA}

\author{A.~Diaz Varela}
\affiliation{Department of Physics, University of Guelph, 50 Stone Road E, Guelph, ON,
N1G 2W1, Canada}

\author{R.~Dunlop}
\affiliation{Department of Physics, University of Guelph, 50 Stone Road E, Guelph, ON,
N1G 2W1, Canada}

\author{L.J.~Evitts}
\altaffiliation[Present address: ]{School of Computer Science and Electronic Engineering, Bangor University, Bangor, Gwynedd, LL57 2DG, UK}
\affiliation{TRIUMF, 4004 Wesbrook Mall, Vancouver, BC, V6T 2A3, Canada}
\affiliation{University of Surrey, Guildford, Surrey GU2 7XH, United Kingdom}

\author{A.B.~Garnsworthy}
\affiliation{TRIUMF, 4004 Wesbrook Mall, Vancouver, BC, V6T 2A3, Canada}

\author{P.E.~Garrett}
\affiliation{Department of Physics, University of Guelph, 50 Stone Road E, Guelph, ON,
N1G 2W1, Canada}

\author{G.~Hackman}
\affiliation{TRIUMF, 4004 Wesbrook Mall, Vancouver, BC, V6T 2A3, Canada}

\author{S.V.~Ilyushkin}
\affiliation{Department of Physics, Colorado School of Mines, 1500 Illinois Street, Golden, CO
80401, USA}

\author{B.~Jigmeddorj}
\affiliation{Department of Physics, University of Guelph, 50 Stone Road E, Guelph, ON,
N1G 2W1, Canada}

\author{K.~Kuhn}
\affiliation{Department of Physics, Colorado School of Mines, 1500 Illinois Street, Golden, CO
80401, USA}

\author{P. Kunz}
\affiliation{TRIUMF, 4004 Wesbrook Mall, Vancouver, BC, V6T 2A3, Canada}
\affiliation{Department of Chemistry, Simon Fraser University, 8888 University Drive, Burnaby BC, V5A 1S6, Canada}

\author{A.T.~Laffoley}
\affiliation{Department of Physics, University of Guelph, 50 Stone Road E, Guelph, ON,
N1G 2W1, Canada}

\author{K.G.~Leach}
\altaffiliation[Present address: ]{Department of Physics, Colorado School of Mines, 1500 Illinois Street, Golden, CO
80401, USA}
\affiliation{TRIUMF, 4004 Wesbrook Mall, Vancouver, BC, V6T 2A3, Canada}

\author{D.~Miller}
\altaffiliation[Present address: ]{Idaho National Laboratory, Idaho
Falls, Idaho 83415, USA}
\affiliation{TRIUMF, 4004 Wesbrook Mall, Vancouver, BC, V6T 2A3, Canada}

\author{W.J.~Mills}
\affiliation{TRIUMF, 4004 Wesbrook Mall, Vancouver, BC, V6T 2A3, Canada}

\author{W.~Moore}
\affiliation{Department of Physics, Colorado School of Mines, 1500 Illinois Street, Golden, CO
80401, USA}

\author{M.~Moukaddam}
\altaffiliation[Present address: ]{PHC-DRS/Universit\'{e} de Strasbourg, IN2P3-CNRS, UMR 7178, F-67037, Strasbourg, France}
\affiliation{TRIUMF, 4004 Wesbrook Mall, Vancouver, BC, V6T 2A3, Canada}

\author{L.N.~Morrison}
\affiliation{TRIUMF, 4004 Wesbrook Mall, Vancouver, BC, V6T 2A3, Canada}
\affiliation{University of Surrey, Guildford, Surrey GU2 7XH, United Kingdom}

\author{B.~Olaizola}
\altaffiliation[Present address: ]{TRIUMF, 4004 Wesbrook Mall,
Vancouver, BC, V6T 2A3, Canada}
\affiliation{Department of Physics, University of Guelph, 50 Stone Road E, Guelph, ON,
N1G 2W1, Canada}

\author{E.E.~Peters}
\affiliation{Department of Chemistry, University of Kentucky, Lexington,
Kentucky 40506-0055, USA}

\author{A.J.~Radich}
\affiliation{Department of Physics, University of Guelph, 50 Stone Road E, Guelph, ON,
N1G 2W1, Canada}

\author{E.T.~Rand}
\affiliation{Department of Physics, University of Guelph, 50 Stone Road E, Guelph, ON,
N1G 2W1, Canada}

\author{F.~Sarazin}
\affiliation{Department of Physics, Colorado School of Mines, 1500 Illinois Street, Golden, CO
80401, USA}

\author{D.~Southall}
\altaffiliation[Present address: ]{Department of Physics, University of Chicago, 5720 South Ellis Avenue, Chicago, IL 60637, USA}
\affiliation{TRIUMF, 4004 Wesbrook Mall, Vancouver, BC, V6T 2A3, Canada}

\author{C.E.~Svensson}
\affiliation{Department of Physics, University of Guelph, 50 Stone Road E, Guelph, ON,
N1G 2W1, Canada}

\author{S.J.~Williams}
\affiliation{National Superconducting Cyclotron Laboratory, Michigan
State University, East Lansing, Michigan 48824, USA}

\author{S.W.~Yates}
\affiliation{Department of Chemistry, University of Kentucky, Lexington,
Kentucky 40506-0055, USA}
\affiliation{Department of Physics \& Astronomy, University of Kentucky,
Lexington, Kentucky 40506-0055, USA}
\date{\today}

\begin{abstract}
We report on high-statistics data from the $\beta^-$ decay of the $^{46}$K $J^{\pi}$ = 2$^-$ ground state taken with the GRIFFIN spectrometer located at the TRIUMF-ISAC facility. In total, 199 $\gamma$ rays and 42 excited states were placed in the level scheme,  and from the observed $\beta$ feeding and angular correlations of pairs of cascading $\gamma$ rays, it was possible to assign spins and parities to  excited states and determine mixing ratios for selected $\gamma$ rays. The level structure of $^{46}$Ca is compared to theoretical predictions from a microscopic valence-space Hamiltonian derived from two- (NN) and three-nucleon (3N) forces. These calculations are in reasonable agreement with the experimental data and indicate that the protons in this region are not as inert as would be expected for semi-magic nuclei. 
\end{abstract}

\maketitle

\section{Introduction}

\begin{figure*}[!ht]
\centering
\includegraphics[scale=0.8]{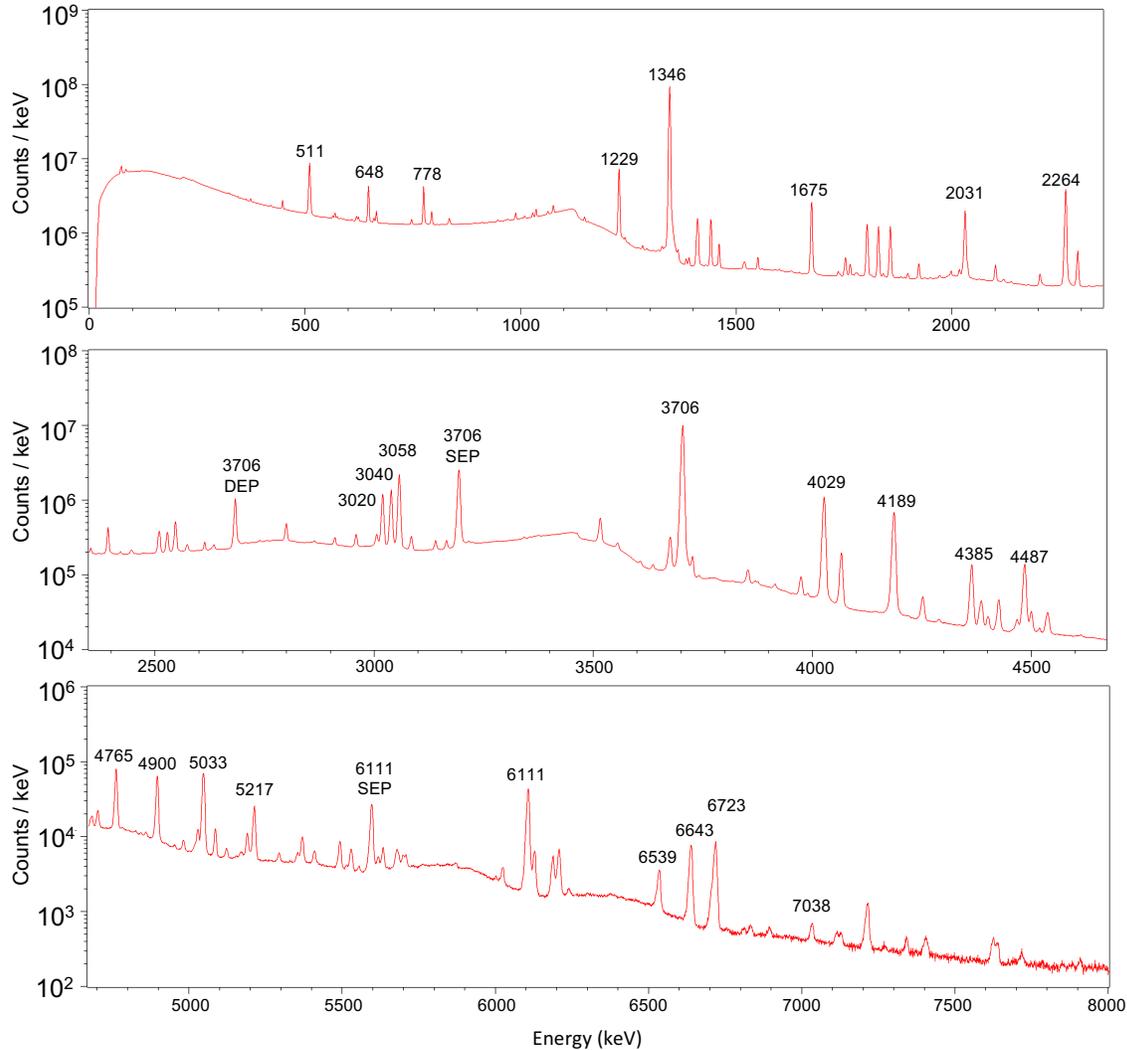}
\vspace*{-7.8cm}
\caption{\label{fig:gammaSing}  $\gamma$-ray singles energy spectrum observed from the $\beta^-$ decay of the $^{46}$K $J^{\pi}$ = 2$^-$ ground state. Selected peaks are labeled with their energies in keV. The abbreviations SEP and DEP indicate single-escape peaks and double-escape peaks, respectively. This spectrum was obtained with 15 HPGe detectors of the GRIFFIN spectrometer in addback mode.}
\end{figure*}

Detailed spectroscopic investigations of calcium and potassium isotopes are critical in understanding the behavior of neutron and proton single-particle orbitals around the $Z$ = 20 shell closure. The structure of these medium-mass nuclei has only recently come within the reach of modern $\textit{ab-initio}$ shell-model calculations that utilize two- (NN) and three-nucleon (3N) forces from chiral effective field theory \cite{jason1, Hebeler, jason2, Soma, Hergert}. To test the finer details of these calculations, spectroscopic information is required, particularly for benchmark regions, such as the calcium isotopes \cite{weinholtz, steppenbeck, ruiz}.

Although $^{46}$Ca is stable, its structure has not been studied in great detail due to its very low natural abundance of 0.004$\%$. Many excited states have been identified in various reactions, most notably from ($p,p'$) \cite{pp}, ($t,p$) \cite{tp}, and ($p,t$) \cite{pt}, but in most cases spin assignments are either tentative or have not been measured \cite{nndc}. The low-lying structure of $^{46}$Ca has been investigated previously in three $\beta$-decay measurements, including two contradictory measurements from the late 1960's by Parsa and Gordon (1966) \cite{parsa} and Yagi. $et~al.$ (1968) \cite{yagi}, and a more recent measurement by Kunz $et~al.$ in 2014 \cite{peter}.

In this article, we present the results of high-statistics data from the $\beta^-$ decay of the $J^{\pi}$ =  2$^-$  $^{46}$K ground state into $^{46}$Ca collected with the GRIFFIN spectrometer (Gamma-Ray Infrastructure For Fundamental Investigations of Nuclei) \cite{griffin, griffin2}, located at TRIUMF-ISAC \cite{TRIUMF}. This measurement expands the current knowledge of the $^{46}$Ca level scheme and includes the placement of $\gamma$ rays from $\gamma$-$\gamma$ coincidence measurements, and spin and parity assignments from the analysis of $\gamma$-$\gamma$ angular correlations in conjunction with calculated log$ft$ values from the observed $\beta$ feeding and decay systematics within the level scheme.

\begin{figure*}[t] 
\centering
\includegraphics[scale=0.8]{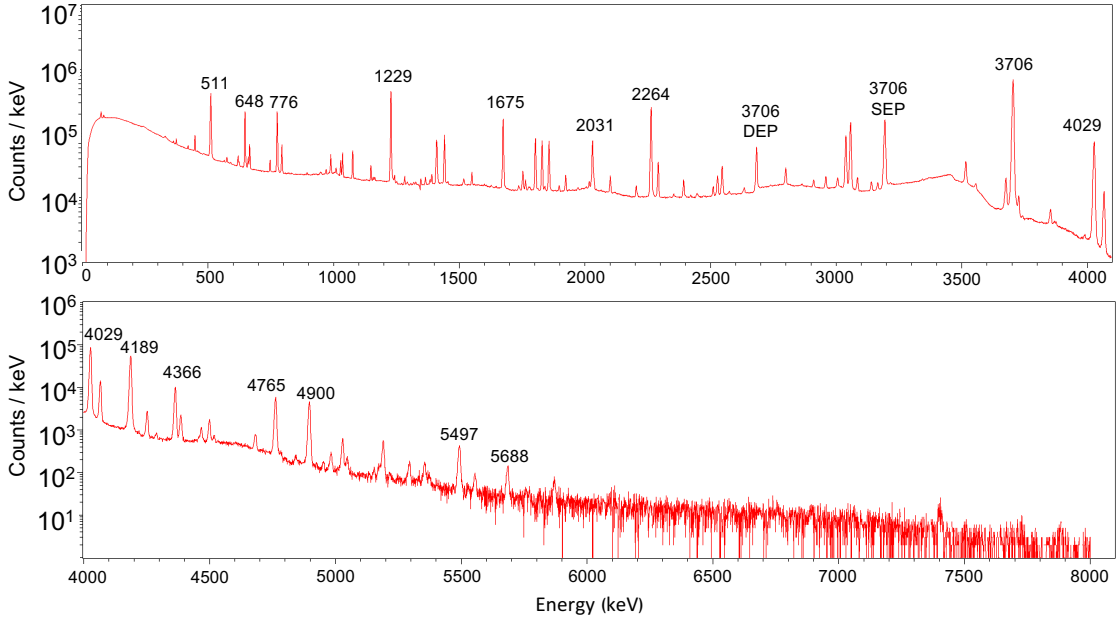}
\vspace*{-13cm}
\caption{\label{fig:gate1345}Spectrum of $\gamma$ rays in coincidence with the 1346 keV 2$_1^+$ $\rightarrow$ 0$_1^+$ $\gamma$ ray. Selected peaks are labeled with their energies in keV.}
\end{figure*}

\section{Experimental Setup}
To produce the $^{46}$K radioactivity, a 9 $\upmu$A 500 MeV proton beam produced by TRIUMF's cyclotron impinged upon a uranium carbide target. The resultant radioactive species were surface ionized and separated with a high-resolution mass separator to select singly charged $A = 46$ ions. A pure beam of 4$\times$10$^5$ pps $^{46}$K was delivered and implanted into Mylar tape at the center of the GRIFFIN array. The array was operated with 15 high-purity germanium (HPGe) clover $\gamma$-ray detectors \cite{usman} (with a HPGe-to-source distance of 11 cm), and was coupled to  PACES \cite{paces, griffin2}, an array of 5 lithium-drifted silicon detectors, operated at liquid-nitrogen temperature, for the detection of conversion electrons placed in the upstream position, and SCEPTAR \cite{sceptar, griffin2}, an array of 10 plastic scintillators for $\beta$-particle tagging, mounted in the downstream position. A sphere of 20 mm thick Delrin plastic absorber was placed around the vacuum chamber to stop energetic $\beta$ particles from reaching the HPGe detectors. For this measurement, the GRIFFIN spectrometer was coupled to an early version of the custom digital acquisition system that is discussed in detail in Ref.~\cite{griffinNIM}.

The experiment was conducted in a cycling mode such that the data were collected in periods of beam-on implantation and beam-off decay. Different cycle times were implemented throughout the experiment, but the beam-on and beam-off collection times were 150 s and 300 s, respectively, for a majority of the experiment. Some data were collected with a longer beam-off period of 1200 s for the purpose of measuring the $^{46}$K half-life. After each decay period, the tape was moved such that the implantation point was subsequently outside of the array and stored behind thick lead shielding to reduce the probability of detection of any long-lived contamination that may have been present. The data were collected in a triggerless singles mode continuously over the course of forty hours.

The efficiencies of the GRIFFIN HPGe detectors from 121 keV to 3.2 MeV were determined from measurements of standard calibration sources, $^{60}$Co, $^{56}$Co, $^{133}$Ba, and $^{152}$Eu, taken at the time of the experiment. The intensities of the observed $\gamma$ rays were corrected for summing effects as is discussed in Ref.~\cite{griffin2}. Several $\gamma$ rays from the decay of $^{46}$K were observed at energies greater than 3.2 MeV, so it was necessary to include simulated high-energy efficiency data points based on GEANT4 simulations that were scaled and fitted to the source data \cite{geant}.

\section{$^{46}$K Decay Scheme}
\begin{figure*}[h!]
\centering
   
		\includegraphics[scale=0.8]{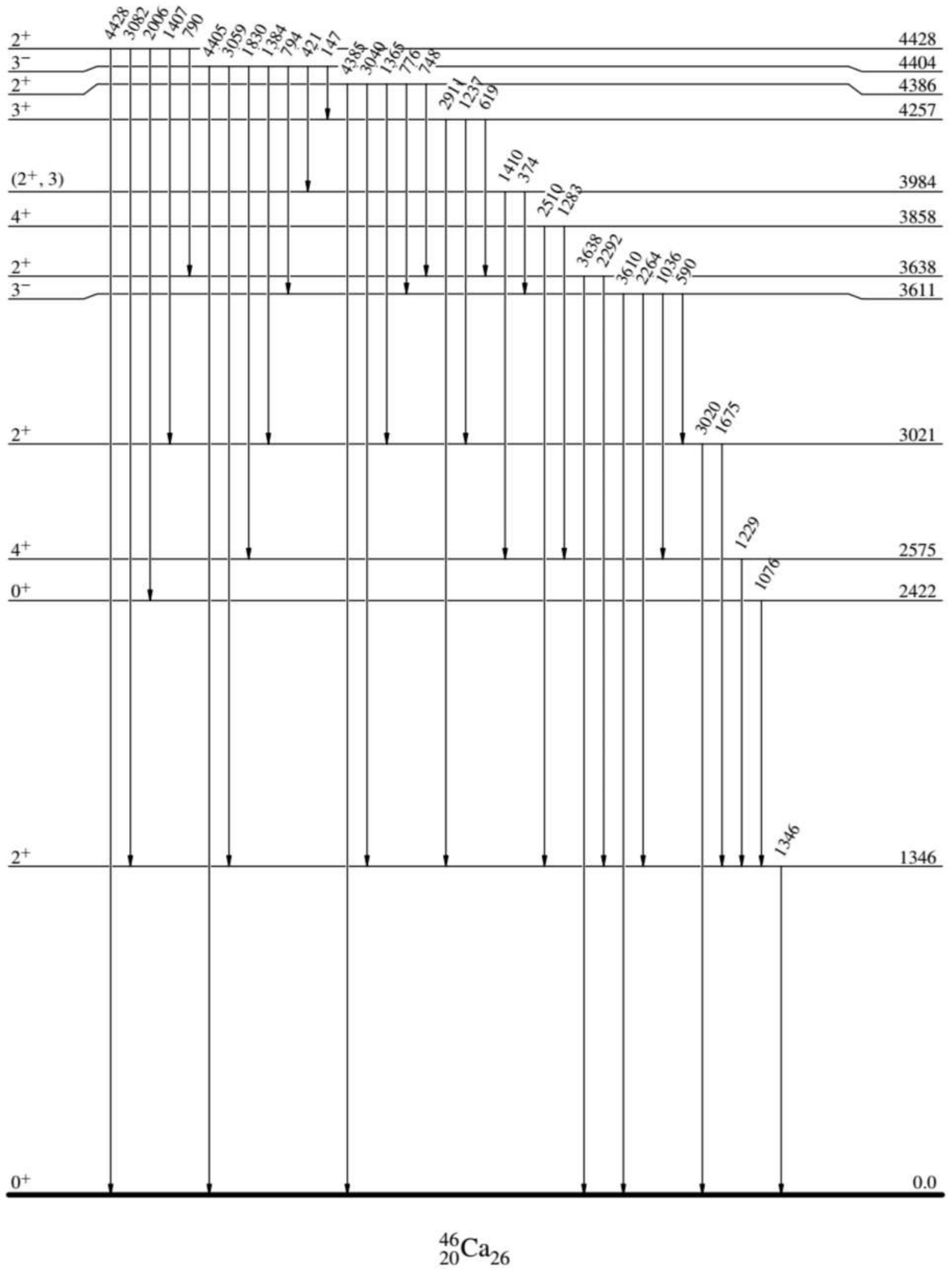}
		\caption{\label{fig:first_levels} A partial level scheme of levels populated in $^{46}$Ca from the $\beta^-$ decay of the $J^{\pi}$ = 2$^- $ $^{46}$K ground state showing $\gamma$ rays that depopulate the 1346-4428  keV levels.}
	
\end{figure*}

\begin{figure*}
\centering
    
		\includegraphics[scale=0.8]{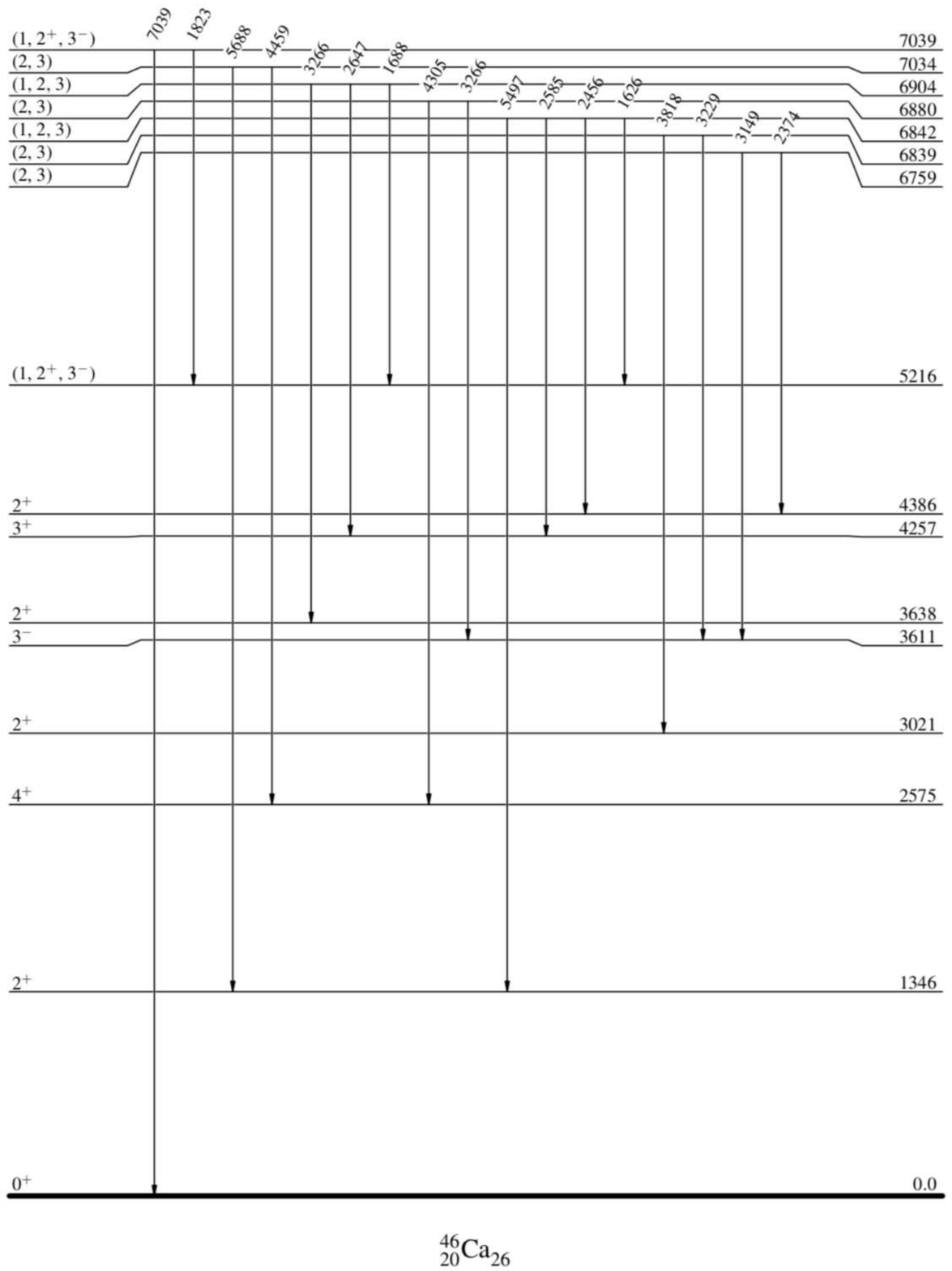}
		\caption{\label{fig:last_levels} A partial level scheme of levels populated in $^{46}$Ca from the $\beta^-$ decay of the $J^{\pi}$ = 2$^- $ $^{46}$K ground state showing $\gamma$ rays that depopulate the 6759-7039 keV levels.}
	
\end{figure*}

From the analysis of these data, the $^{46}$K decay scheme has been greatly expanded to contain 199 $\gamma$ rays  and 42 excited states in $^{46}$Ca. The complete level scheme is shown in Figures~\ref{fig:first_levels} - \ref{fig:last_levels} and the observed energies, intensities, and branching ratios of the $\gamma$ rays along with the energies and possible $J^{\pi}$ assignments of the excited states are listed in Table \ref{tab:bigTable}.

\subsection{$\gamma$ Singles Events}
The analysis was performed using the addback capabilities of the GRIFFIN HPGe clover detectors, such that all energies detected within a clover detector in a 250 ns coincidence-timing window were summed together to minimize the number of Compton-scattering events observed \cite{usman}. Roughly 9$\times$10$^8$ $\gamma$-ray singles events were observed; see  Fig.~\ref{fig:gammaSing} for the $\gamma$-singles energy spectrum. For $\gamma$-ray intensities obtained from the singles events, the fitted peak areas were corrected for summing effects using the empirical method described in Ref.~\cite{griffin2}.

\subsection{$\gamma$ - $\gamma$ Coincidences}
To place $\gamma$ rays in the $^{46}$Ca level scheme, $\gamma$-$\gamma$ coincidences were analyzed, where $\gamma$ rays were considered in coincidence if they were detected within a 250 ns window. Additionally, the coincidence events were time-random-background subtracted to eliminate false coincidences. An energy-coincidence gate taken on the 2$_1^+$ $\rightarrow$ 0$_1^+$ 1346 keV $\gamma$ ray is presented in Fig.~\ref{fig:gate1345} and illustrates the quality of the data.

%%%%%%%%%%%%%Big Table%%%%%%%%%%%%
\clearpage
\begin{longtable*}{|cccccc|}
\caption{The $^{46}$Ca levels populated in the $\beta^-$ decay of the $^{46}$K 2$^-$ ground state. These are presented by parent level, where the intensity ($I_{\gamma}$) of each $\gamma$ ray has been determined relative to that of the 1346 keV $\gamma$ ray.  The $\gamma$-ray branching ratio ($BR_\gamma$) of a given $\gamma$ ray is the ratio of the intensity of that $\gamma$ ray relative to the intensity of the strongest $\gamma$ ray that depopulates the same level. Spin-parity assignments ($J^{\pi}$) that are the result of this work have been specified, otherwise they are rom Ref. \cite{nndc}.} \label{tab:bigTable}\\

\hline\hline
\multicolumn{1}{|c}{$E_{inital}$ (keV)} &$J^{\pi}$&\multicolumn{1}{c}{$E_{\gamma}$ (keV)}&\multicolumn{1}{c}{ $E_{final}$ (keV)}&\multicolumn{1}{c}{$I_{\gamma}$}&\multicolumn{1}{c|}{$BR_{\gamma}$}\\ \hline
\endfirsthead

\multicolumn{3}{c}
{{\tablename\ \thetable{} -- continued from previous page}} \\
\hline \multicolumn{1}{|c}{$E_{inital}$ (keV)} &$J^{\pi}$&\multicolumn{1}{c}{$E_{\gamma}$ (keV)}&\multicolumn{1}{c}{ $E_{final}$ (keV)}&\multicolumn{1}{c}{$I_{\gamma}$}&\multicolumn{1}{c|}{$BR_{\gamma}$}\\ \hline
\endhead

\hline \multicolumn{6}{r}{{Continued on the next page}}\\ 
\endfoot

\hline \multicolumn{6}{l}{{$\dag$ $J^{\pi}$ assignment or constraint is the result of the $\gamma$-$\gamma$ angular correlation analysis in conjunction with calculated}}\\ 
\multicolumn{6}{l}{{\indent log$ft$ and decay systematics (when possible).}}\\
\multicolumn{6}{l}{{$\ddag$ $J^{\pi}$ assignment constrained from calculated log$ft$ and decay systematics (when possible).}}
\endlastfoot

1346.00(14) & 2$^+$ & 1345.96(26) & 0.00 & 100 & 100\\
\hline
2421.98(22) & 0$^+$ & 1076.18(26) & 1346.00 & 0.441(8) & 100\\
\hline
2574.53(16)& 4$^+$ & 1228.72(26) & 1346.00 & 6.52(12) & 100\\
\hline
3020.67(14) & 2$^+$ & 1675.09(26) & 1346.00 & 3.20(6)  & 100 \\
       &         & 3020.21(26) & 0.00    & 2.43(6)  &76.0(24)\\
\hline
3610.51(15)& 3$^-$ & 590.27(26) & 3020.67 & 0.0127(8) & 0.188(12)\\
       &         & 1036.16(26)& 2574.53 & 0.418(7)  & 6.21(16)\\
       &         & 2264.29(26)& 1346.00 & 6.72(13)  & 100\\
       &         & 3610.05(84)& 0.00    & 0.0635(10)  & 0.944(154)\\       
\hline
3638.38(15) & 2$^+$ & 2292.15(26) & 1346.00 & 0.679(14) & 100\\
       &         & 3638.39(84) & 0.00    & 0.0486(17)& 7.16(30)\\
\hline
3857.51(17) & 4$^+_{\dag}$& 1283.11(26) & 2574.53 & 0.132(3) & 100\\
     &         & 2509.92(26) & 1346.00 & 0.0766(30)& 88.3(38)\\
\hline
3984.07(16) &(2$^+$, 3)$_{\ddag}$ & 374.11(26) & 3610.51 & 0.137(4) & 26.9(15)\\
     &         & 1409.81(26) & 2574.53 & 0.511(22)& 100\\
\hline
4257.26(15) &  3$^+_{\dag}$   & 619.26(27) & 3638.38 & 0.00896(49) & 5.27(34)\\
       &         & 1236.50(30)& 3020.67 & 0.0101(10)  & 8.14(82)\\
       &         & 2911.08(26)& 1346.00 & 0.167(6)  & 100\\
\hline
4385.91(15) & 2$^+_{\dag}$       & 747.73(26) & 3638.38& 0.180(7) & 5.69(26)\\
       &         & 775.64(26) & 3610.51& 2.44(8)  & 76.5(32)\\
       &         & 1365.47(26)& 3020.67 & 0.123(3)  & 5.36(20)\\
       &         & 3040.37(83)& 1346.00 & 3.07(9)  & 100\\   
       &         & 4385.30(83)& 0.00    & 0.116(4)  & 3.67(17)\\    
\hline
4404.48(15) & 3$^-_{\dag}$  & 147.22(21) & 4257.26 & 0.00537(4)& 0.0955(79) \\
       &         & 420.99(26)& 3984.07& 0.0228(8)  & 0.405(18)\\
       &         & 794.09(26)& 3610.51 & 0.398(13)  &7.08(30)\\
       &         & 1384.15(26)& 3020.67 & 0.0983(25)  &1.75(7) \\   
       &         & 1830.10(26)& 2574.53    &1.55(4) &27.6(10) \\       
       &         & 3058.83(83)& 1346.00   & 5.62(16)&100 \\  
       &         & 4405.15(83)& 0.00   & 0.0232(10)&0.41(2) \\  
\hline
4428.02(14) & 2$^+_{\dag}$           & 789.55(26) & 3638.38& 0.0191(8) & 14.8(9)\\
       &         & 1407.45(26) &3020.67 & 0.126(6)  & 97.5(62)\\
       &         & 2006.26(26)& 2421.98 & 0.0268(8)  & 20.7(10)\\
       &         & 3082.04(20)& 1346.00 & 0.0784(30)  & 61(23)\\   
       &         & 4428.08(83)& 0.00   & 0.129(5)  & 100\\       
\hline
4432.05(16) &  (3, 4$^+$)$_{\dag}$   & 448.65(26) & 3984.07 & 0.419(12) & 28.3(10)\\
       &          & 574.40(26) &3857.51 & 0.136(7)  & 9.17(51)\\
       &          & 793.87(26) & 3638.38 & 0.0541(21)  & 3.65(16)\\
       &          & 1411.383(22)  & 3020.67 & 0.661(20) & 44.6(15)\\   
       &          & 1857.72(26)& 2574.53& 1.48(3)  & 100\\       
       &          & 3086.34(83)& 1346.00 & 0.271(7)  & 18.3(6)\\
\hline
4487.05(16) & 2$^+_{\dag}$    & 3141.52(83) & 1346.00 & 0.204(7) & 37.3(19)\\
     &         & 4487.37(83) & 0.00 & 0.548(21)& 100\\
\hline
4991.83(21) &(1, 2, 3, 4$^+$)$_{\ddag}$ & 1971.12(27) & 3020.67 & 0.0258(11) & 100\\
\hline
5051.96(16) &  2$^-_{\dag}$ & 565.01(26) & 4487.05 & 0.0948(19) &0.266(9)\\
       &          & 619.95(26)   &4432.06& 0.162(7)  & 0.454(23)\\
       &          & 624.33(26)   &4428.02 & 0.184(3)  & 0.516(17)\\
       &          & 647.67(26)   &4404.48 & 1.97(8) & 5.51(27)\\   
       &          & 666.14(26)   &4385.91 & 0.451(7)  & 1.26(4)\\       
       &          & 1413.68(26)& 3638.38 & 0.105(4)  & 0.294(14)\\
       &          & 1441.64(26)& 3610.51 & 1.39(4)  & 3.89(17)\\
       &          & 2031.01(26)& 3020.67 & 2.87(7)  & 8.05(30)\\
       &          &3705.94(83) & 1346.00 & 35.7(10)  & 100\\
      
\hline
5216.36(17) &(1, 2$^+$, 3$^-$)$_{\ddag}$   & 5217.01(83) & 0.00 & 0.120(6) & 100\\
\hline
5351.19(18) &  (4$^+$)$_{\ddag}$    & 919.24(26) & 4432.06 & 0.00723(90) & 26.6(35)\\
       &         & 1367.41(27)& 3984.07 & 0.0115(8)  & 42.2(33)\\
       &         & 1740.71(30)& 3610.51 & 0.0272(12)  & 100\\
       &         & 2776.41(40)& 2574.53 & 0.00485(47)  & 17.9(19)\\       
\hline
5374.97(16) &  3$^-_{\dag}$  & 947.12(26) & 4428.02 & 0.138(7) &3.15(16)\\
       &          & 970.11(26)   &4404.48 & 0.0644(4)  & 1.47(7)\\
       &          & 988.99(26)   &4385.91 & 0.325(13)  & 7.40(30)\\
       &          & 1390.95(26)  &3984.07 & 0.138(4) & 3.14(10)\\   
       &          & 1519.19(26)  &3857.51 & 0.119(6)  & 2.71(15)\\       
       &          & 1736.85(26)  &3638.38 & 0.0559(22)  & 1.27(5)\\
       &          & 1764.47(26)  &3610.51 & 0.121(4)  & 2.77(10)\\
       &          & 2353.92(26)& 3020.67& 0.0610(41)  & 1.39(9)\\
       &          & 2800.16(26) & 2574.53 & 0.517(2)  & 11.8(1)\\
       &          & 4028.95(83) & 1346.00 & 0.439(14)  & 100\\
        &          & 5374.47(83) & 0.00 & 0.0316(18)  & 0.721(40)\\
\hline
5414.42(15) &(1, 3)$^-_{\dag}$     & 197.97(26) & 5216.36 & 0.00576(24) &0.350(19)\\
       &          & 362.98(26)   &5051.96 & 0.0467(27)  & 2.81(19)\\
       &          & 927.07(26)   &4487.05 & 0.0320(17)  & 1.95(12)\\
       &          & 982.52(29) & 4432.06 & 0.0130(13) & 0.79(8)\\
       &          & 986.48(27)  &4428.02 & 0.0154(1) & 0.934(68)\\   
       &          & 1009.94(26)  &4404.48 & 0.0570(28)  & 3.47(21)\\       
       &          & 1028.44(26)  &4385.91 & 0.231(9)  & 14.0(7)\\
       &          & 1157.10(26)  &4257.26 & 0.0385(22)  & 2.32(16)\\
       &          & 1775.99(26)& 3638.38& 0.0256(11)  & 1.56(8)\\
       &          & 1803.95(26) & 3610.51 & 1.64(5)  & 100\\
       &          & 2393.38(26) & 3020.67 & 0.451(23)  & 27.4(11)\\
       &          & 4068.51(83) & 1346.00 & 0.658(23)  & 40.1(20)\\
       &          & 5414.21(84) & 0.00 & 0.0137(1)  & 0.834(6)\\
\hline
5534.53(17) &  (2$^+$, 3$^-$)$_{\dag}$        & 1106.57(26) & 4428.02 & 0.0234(14) &0.656(46)\\
       &          & 1148.60(26)   &4387.05 & 0.200(8)  & 5.61(31)\\
       &          & 1550.54(26)   &3984.07 & 0.212(7)  & 5.94(28)\\
       &          & 1677.15(26)  &3857.51 & 0.0410(24) & 1.15(8)\\   
       &          & 1923.92(26)  &3610.51 & 0.237(8)  & 6.65(34)\\       
       &          & 2511.09(27)  &3020.67 & 0.0261(9)  & 0.732(38)\\
       &          & 2959.57(28)  &2574.52 & 0.313(9)  & 8.79(41)\\
       &          & 4188.58(83)& 1346.00& 3.56(13)  & 100\\
       &          & 5536.29(98)& 0.00 & 0.0219(15)  & 0.615(49)\\
\hline
5712.24(16) &    (2, 3$^+$)$_{\dag}$        & 495.89(26) & 5216.36 & 0.00574(25) &0.853(50)\\
       &          & 660.41(26)   &5051.96 & 0.172(18)  & 25.6(15)\\
       &          & 1225.04(27)   &4487.05 & 0.0142(8) & 2.11(14)\\
       &          & 1307.46(30)  &4404.48 & 0.0130(11) & 1.94(18)\\   
       &          & 1326.56(28)  &4385.91 & 0.0170(12)  & 2.53(20)\\       
       &          & 1454.98(26)  &4257.26 & 0.0281(17)  & 4.18(37)\\
       &          & 2073.55(30)  &3638.38 & 0.00277(22)  & 0.412(37)\\
       &          & 2101.51(26)& 3610.51 & 0.239(8)  & 35.5(19)\\
       &          & 2691.79(26)& 3020.67 & 0.0261(10)  & 3.87(21)\\
       &          & 4366.27(83)& 1346.00 & 0.673(26)  & 100\\
\hline
5815.36(17) &   (1, 2, 3)$_{\ddag}$   & 599.25(37) & 5216.36 &  0.00073(10)&0.538(70)\\ 
       &            & 1328.32(26) & 4487.05 & 0.0851(46)  & 63.0(40)\\
       &          & 1411.21(26)   &4404.48 & 0.0793(37)& 58.6(34)\\
       &          & 1429.68(27)   &4385.91 & 0.0106(64)& 7.83(55)\\
       &          & 1830.12(27)  &3984.07& 0.0219(9) & 16.2(84)\\   
       &          & 2176.88(27)  &3638.38 & 0.00154(18)  & 1.14(14)\\       
       &          & 2204.72(26)  &3610.51 & 0.135(5)  & 100\\
       &          & 2795.21(32)  &3020.67 & 0.00182(18)  & 1.35(14)\\
       &          & 4470.30(84)& 1346.00 & 0.0392(19)  & 29.0(18)\\

\hline
5848.20(18) & (1, 2, 3)$_{\ddag}$          & 1443.95(27) & 4404.48 & 0.0182(10)&21.7(15)\\
       &          & 1591.20(27)   &4257.26 & 0.00479(34)  & 5.72(48)\\
       &          & 1863.83(27)  &3984.07 & 0.0134(6)& 16.0(10)\\
       &          & 1990.56(29)  &3857.51 & 0.0277(12)&33.2(20) \\   
       &          & 2209.54(26)  &3638.38 & 0.00208(19) & 2.49(25)\\       
       &          & 4502.19(83) &1346.00 & 0.0836(37)  & 100\\
\hline
6032.21(17) &  (2$^+$, 3$^-$)$_{\ddag}$         & 815.51(27) & 5216.36 & 0.00514(24)&13.3(9)\\
       &          & 1600.25(26)   &4432.06 & 0.0244(12)  & 63.1(45)\\
       &          & 1627.98(27)  &4404.48& 0.0132(8)& 34.1(26)\\
       &          & 1646.32(26)  &4385.91 & 0.0159(8)&41.2(30) \\   
       &          & 2421.56(27)  &3610.51 & 0.0231(9) & 59.7(39)\\  
        &          & 3012.11(84) &3020.67 & 0.00562(32)  & 14.5(10)\\    
       &          & 4685.40(84) &1346.00 & 0.0387(20)  & 100\\
       &          & 6031.1(84) &0.00& 0.0101(7)  & 26.1(23)\\
\hline
6111.14(28) &  (2$^+$, 3$^-$)$_{\dag}$    & 1624.14(28) & 4487.05& 0.00407(22) & 0.879(62)\\
       &         & 1681.80(75) & 4428.02 & 0.00557(80)  & 1.20(18)\\
       &         & 4765.36(83)& 1346.00 & 0.463(21)  & 100\\
       &         & 6111.13(83)& 0.00 & 0.341(24) & 73.6(62)\\       
\hline
6245.43(16) &   (2, 3)$_{\ddag}$       & 894.52(26) & 5351.19 & 0.0189(12)&4.94(40)\\
       &          & 1253.72(27)   &4991.83 & 0.00244(14)  & 0.640(49)\\
       &          & 1813.54(29)   &4432.05 & 0.0122(11)& 3.19(34)\\
       &          & 1840.99(26) &4404.48& 0.0474(23) & 12.4(84)\\   
       &          & 1859.55(26) &4385.91 & 0.0195(10)  & 5.11(35)\\       
       &          & 1987.97(27)  &4257.26 & 0.00572(40) & 1.50(13)\\
       &          & 2261.27(26)  &3984.07 & 0.115(38)  & 30.1(17)\\
       &          & 2635.09(26)& 3610.51 & 0.0773(28) & 20.3(12)\\
       &          &3224.47(84) &3020.67 & 0.0202(8)  & 5.30(33)\\
       &          & 3671.00(83)  &2574.53 & 0.0230(8)  & 6.03(37)\\
       &          & 4899.74(83)& 1346.00 & 0.382(18)  & 100\\
\hline
6302.67(16) & (2, 3)$_{\ddag}$      & 951.51(26) & 5351.19& 0.0184(12)&6.44(47)\\
       &          & 1815.50(32)   &4487.06& 0.00240(13)  & 0.84(5)\\
       &          & 1898.23(26)  &4404.48 & 0.0681(32)& 23.9(14)\\
       &          & 1916.70(27) &4385.91 & 0.00863(52) & 3.03(21)\\   
       &          & 2045.36(27) &4257.26 & 0.00699(48)  & 2.45(18)\\       
       &          & 2319.12(30) &3984.07 & 0.00579(55) & 2.03(20)\\
       &          & 2444.83(27)  &3857.51 & 0.0493(29)  & 17.3(12)\\
       &          & 2691.79(26)& 3610.51& 0.0133(6) & 4.66(26)\\
       &          & 3281.92(83) &3020.67 & 0.0111(6) & 3.89(23)\\
       &          & 3727.72(83)  &2574.53 & 0.285(9)  & 100\\
\hline
6377.80(18)&  (1, 2, 3)$_{\ddag}$       & 1385.81(27) & 4991.83 & 0.00756(24)&16.7(10)\\
       &          & 1890.74(65)   &4487.05 & 0.0167(9)  & 36.8(28)\\
       &          & 1950.04(27)  &4428.02 & 0.0197(13)& 43.4(37)\\
       &          & 1973.94(26)&4404.48 & 0.0133(8) & 29.3(23)\\   
       &          & 2120.62(26) &4257.26& 0.00256(20) & 5.65(54)\\       
       &          & 2739.17(26) &3638.38 & 0.0301(13)& 66.2(46)\\
       &          & 3356.73(83)  &3020.67 & 0.0122(5)  & 26.8(19)\\
       &          & 5032.52(83)& 1346.00 & 0.0454(24) & 100\\
\hline
6503.00(18)&   (1, 2, 3)$_{\ddag}$        & 1286.37(27) & 5216.36 & 0.00428(25)&12.3(9)\\
       &          & 2074.93(30)   &4428.02 & 0.00708(61)  & 20.3(20)\\
       &          & 2098.66(32)  &4404.48& 0.00324(27)& 9.29(91)\\
       &          & 2117.49(28)&4385.91 & 0.00289(29) & 8.27(91)\\   
       &          & 2864.24(26) &3638.38 & 0.0349(17) & 100\\       
\hline
6540.09(60)&  (1, 2$^+$, 3$^-$)$_{\ddag}$        & 5193.96(83) & 1346.00 & 0.0374(22)&100\\
       &          &6539.40(84) &0.00 & 0.0234(20)  & 62.5(64)\\
\hline
6565.62(19) & (2, 3)$_{\ddag}$        & 2137.70(27) & 4428.02 & 0.0203(13)&100\\
       &          & 2161.41(28)  &4404.48 & 0.00631(42)  & 31.1(29)\\
       &          & 2926.71(29)  &3638.38 & 0.00477(30)& 23.5(21)\\
       &          & 2954.71(29)&3610.51 & 0.00948(50) & 46.8(39)\\   
\hline
6643.37(23)&   (1, 2$^+$, 3$^-$)$_{\ddag}$        & 2155.46(29) & 4487.05 & 0.00486(27)&24.0(25)\\
       &          & 2215.09(28)  &4428.02 & 0.00427(47)  & 6.57(92)\\
       &          & 4220.56(83) &2421.98 & 0.00791(40)& 12.2(12)\\
       &          & 5297.44(83)&1346.00 & 0.0118(8) & 18.1(20)\\   
       &          & 6643.25(83)&0.00 & 0.0650(56) & 100\\ 
\hline
6704.98(84)&   (1, 2, 3)$_{\ddag}$       & 5358.64(83) & 1346.00 & 0.0107(8)&100\\
\hline
6722.14(21) &  (2$^+$, 3$^-$)$_{\ddag}$            & 1506.64(28) &5216.363 & 0.00271(19)&3.40(39)\\
       &          & 2294.86(32) &4428.02 & 0.00427(47) & 5.37(76)\\
       &          & 2335.64(30) &4385.91 & 0.00824(52)& 10.3(11)\\
       &          & 3701.80(83) &3020.67& 0.0103(5) & 12.9(13)\\   
       &          & 4147.07(83) &2574.53 & 0.00546(34) & 6.87(75)\\ 
       &          & 5375.79(83) &1346.00& 0.0316(17) & 39.8(42)\\ 
       &          & 6722.95(83) &0.00 & 0.0795(71) & 100\\ 
\hline
6759.22(25) &  (2, 3)$_{\ddag}$          & 2373.51(33) & 4385.91 & 0.00181(22)&12.4(16)\\
       &          &3149.19(84)  &3610.51 & 0.0146(7)  & 100\\
\hline
6838.75(30) &  (2, 3)$_{\ddag}$     &1687.76(40)&5216.36  & 0.00140(14)&11.2(12)\\     
       &          & 3228.91(83) &3610.51 & 0.0125(6)& 100\\
       &          & 3818.37(84)&3020.67 & 0.00210(18) & 16.8(16)\\   
    
\hline
6842.11(24)    &  (1, 2, 3)$_{\ddag}$           &1625.97(43) & 5216.36 & 0.000820(140) & 1.80(32)\\
                &          &2456.00(30) & 4385.91 & 0.00344(30) & 7.57(81)\\
                &          & 2584.64(36)  &4257.26& 0.00103(12) & 2.27(30)\\
                &          & 5496.58(83)&1346.00 & 0.0455(28) & 100\\ 
\hline
6879.91(60) &   (2, 3)$_{\ddag}$         & 3265.86(83) & 3610.51& 0.00166(16)&46.8(58)\\
      &          &4304.87(83)  &2574.53 & 0.00355(27)  & 100\\

\hline
6904.19(26) &  (1, 2, 3)$_{\ddag}$          & 1687.60(31) & 5216.36 & 0.00140(14)&23.7(27)\\
       &          &2647.04(32)  &4257.26 & 0.00188(18)  & 31.9(35)\\
       &          &3265.86(83) &3638.38 & 0.00589(36)  & 100\\
\hline
7033.64(29) &  (2, 3)$_{\ddag}$           &4459.06(84) &2574.53 & 0.00378(36)  & 63.2(112)\\
   &          & 5688.09(84) & 1346.00 & 0.00598(36)&100\\
\hline
7039.25(37)&  (1, 2$^+$, 3$^-$)$_{\ddag}$            & 1822.82(36) & 5216.36& 0.00105(13)&34.0(55)\\
    
       &          &7038.83(92)  &0.00 & 0.00309(34)  & 100\\
\end{longtable*}
%%%%%%%%%%%%%%%%%%%%%%%%%%%%%%%%%%%%%%%%%%%%%%%%%

When it was not possible to directly obtain the $\gamma$-ray intensity from the singles spectrum, the intensity was obtained by gating from below the $\gamma$ ray of interest, given that
\begin{equation}
N_{12}=NI_{\gamma1}\epsilon(\gamma_{1})B_{\gamma2}\epsilon(\gamma_{2})\epsilon_{12}\eta(\theta_{12}), 
\end{equation}
where $N_{12}$ is the number of counts in a coincidence peak between two cascading $\gamma$ rays, $N$ is the overall normalization factor that characterizes the specific decay data set, $I_{\gamma1}$ is the intensity of the ``feeding'' $\gamma$ ray ($\gamma_{1}$), $B_{\gamma2}$ is the branching ratio of $\gamma_2$ relative to all intensity depopulating the same excited state, $\epsilon(\gamma_{1})$ and $\epsilon(\gamma_{2})$ are the singles relative photopeak efficiencies, $\epsilon_{12}$ is the coincidence efficiency, and $\eta$($\theta_{12}$) is the angular correlation factors described in Ref. \cite{kulp}.

 Note that it was important to correct $N_{12}$ for possible summing effects.  This has been done here for coincidence data using a similar method to that discussed for singles data in Ref \cite{griffin2}. In the coincidence case, the summing correction factors are specific to the transition of interest as well as the choice of the gating transition. Care must be taken when constructing the necessary coincidence matrices used to determine these factors so that the same experimental conditions are applied to them as to the experimental data.

\begin{table}
\centering
\caption{\label{tab:betaFeeding}The $\beta$ feeding observed compared with the two previous measurements. In total, 42 excited states were observed and 40 of them are populated by the $\beta^-$ decay of the $^{46}$K 2$^-$ ground state ($Q_{\beta}$ = 7.7716(16) MeV \cite{nndc} $T_{1/2}$ = 96.303(79) s  \cite{peter}). Lower limits are given for states that were not observed to be populated.}

\begin{tabular}{cccl|cc}
\hline
\hline
$E_{level}$ (keV)&$\%$&&log$ft$&$\%$\cite{parsa}&\%\cite{yagi}\\
\hline
1346.00(14)&27.4(21)&&6.92(4)&50&63\\
2421.98(22)&0.38(1)&&10.27(2)$^{1u}$&&\\
2574.53(16)&1.15(14)&&9.70(6)$^{1u}$&&3\\
3020.67(14)&0.55(20)&&8.01(16)&11&\\
3610.51(15)&0.28(14)&&8.04(22)&8&5\\
3638.38(15)&0.22(2)&&8.13(4)&&\\
3857.51(17)&$\textless$0.003&&$\textgreater$11.5$^{1u}$&&\\
3984.07(16)&0.47(3)&&7.63(3)&&\\
4257.26(15)&0.092(9)&&8.20(5)&&\\
4385.91(15)&4.39(16)&&6.44(2)&&\\
4404.48(15)&5.09(22)&&6.37(2)&&\\
4428.02(14)&0.017(14)&&8.80(40)&&\\
4432.05(16)&2.59(7)&&6.65(2)&&\\
4487.05(16)&0.46(2)&&7.36(2)&3&0.6\\
4991.83(16)&0.015(1)&&8.85(3)&&\\
5051.96(16)&39.7(13)&&5.06(2)&28&25\\
5216.36(17)&0.086(6)&&7.61(4)&&\\
5351.19(18)&$\textless$0.005&&$\textgreater$9.96$^{1u}$&&\\
5374.97(16)&5.52(13)&&5.68(2)&&\\
5414.42(15)&2.99(9)&&5.91(2)&&\\
5534.53(17)&4.29(16)&&5.66(2)&&\\
5712.24(16)&1.10(4)&&6.09(2)&&\\
5815.26(17)&0.35(1)&&6.49(2)&&\\
5848.20(18)&0.14(1)&&6.86(4)&&\\
6032.21(17)&0.13(1)&&6.71(2)&&\\
6111.14(28)&0.75(3)&&5.86(3)&&\\
6245.43(16)&0.67(2)&&6.61(4)&&\\
6302.67(16)&0.43(2)&&5.88(3)&&1.5\\
6377.80(18)&0.14(1)&&6.27(4)&&\\
6503.00(18)&0.049(2)&&6.56(3)&&\\
6540.09(60)&0.056(3)&&6.44(4)&&\\
6565.62(19)&0.038(2)&&6.57(4)&&\\
6643.37(23)&0.087(6)&&6.10(4)&&2.1\\
6704.98(84)&0.010(8)&&6.94(5)&&\\
6722.14(21)&0.13(1)&&5.79(5)&&\\
6759.22(25)&0.015(1)&&6.67(4)&&\\
6838.75(30)&0.015(1)&&6.52(5)&&\\
6842.11(24)&0.047(3)&&6.02(5)&&\\
6879.91(60)&0.0048(3)&&6.94(5)&&\\
6904.19(26)&0.0086(5)&&6.64(5)&&\\
7033.64(29)&0.0090(9)&&6.34(6)&&\\
7039.25(37)&0.0038(4)&&6.70(6)&&\\
\hline \multicolumn{6}{l}{{$^{1u}$log$ft$ calculated for a unique first-forbidden transition.}}\\
\end{tabular}

\end{table}

\subsection{$\beta$ Feeding}
The $Q_{\beta}$ value for the decay of $^{46}$K 2$^-$ ground state to excited states in $^{46}$Ca is 7.7716(16) MeV \cite{nndc}. In the analysis of these data, excited states were placed in the $^{46}$Ca level scheme with energies up to approximately 7039 keV. To determine the $\beta$ feeding to these states, the $\gamma$-ray intensity balance of each excited state was investigated. Any unobserved intensity populating an excited state was attributed to $\beta$ feeding from the parent nucleus, such that the total missing intensity populating all of the excited states is equivalent to the total $\beta$-feeding intensity. The $\beta$-feeding branching ratios and log$ft$ values are presented in Table~\ref{tab:betaFeeding} for each excited state. Note that  these $\beta$-feeding percentages were determined assuming no $\beta$ feeding to the $^{46}$Ca ground state from the $^{46}$K parent.  Previously, Parsa and Gordon \cite{parsa} measured the $\beta$ continuum for this decay and had not observed any intensity above 6.5 MeV. A decay to the $^{46}$Ca ground state would be a unique first-forbidden transition, assuming log$ft$ = 10 (approximately what was observed for the $^{44}$K 2$^-$ ground state to $^{44}$Ca 0$^+$ ground state decay \cite{44K}) then a potential upper limit of intensity for such a branch is 8\%.

\begin{figure}
\centering
\vspace*{-1.2cm}
\hspace*{-1cm}
\includegraphics[scale=0.55]{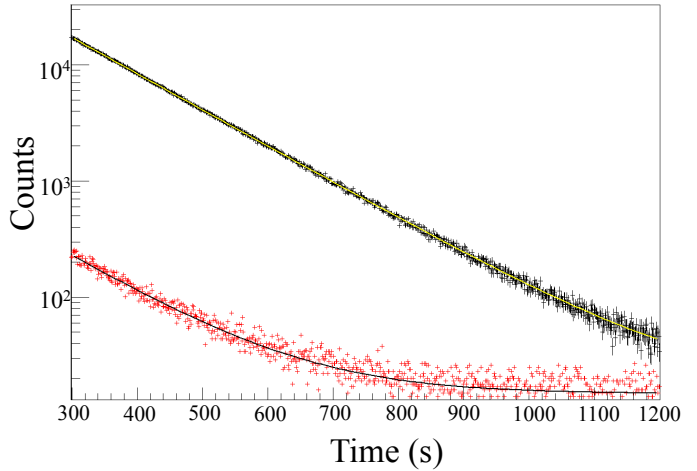}
\vspace*{-4.7cm}
\caption{\label{fig:halflife} The decay of the 1346 keV $\gamma$ ray (shown in black) was fit to determine the $T_{1/2}$ for the $\beta^-$ decay of the $^{46}$K 2$^-$ ground state (fit shown in yellow) taking into account the Compton-scattering background (data shown in red with respective fit in black). The fit $T_{1/2}$ from the decay of this $\gamma$ ray is 96.41(10)s.}
\end{figure}

\section{$^{46}$K Half-Life Measurement}
Three studies have reported a half-life ($T_{1/2}$) measurement for the $\beta^-$ decay of $^{46}$K. Currently the Nuclear Data Sheets for $A$ = 46 \cite{nndc} report $T_{1/2}$ = 105(10) s as a weighted average of the two earliest measurements of 115(4) s from Parsa and Gordon \cite{parsa} and 95(5) s from Yagi $et~al.$ \cite{yagi}, and does not take into account the recent, more precise measurement of 96.303(79) s from Kunz $et~al.$ \cite{peter}.

\begin{figure*}
\centering
\hspace*{1cm}
\includegraphics[scale=0.65]{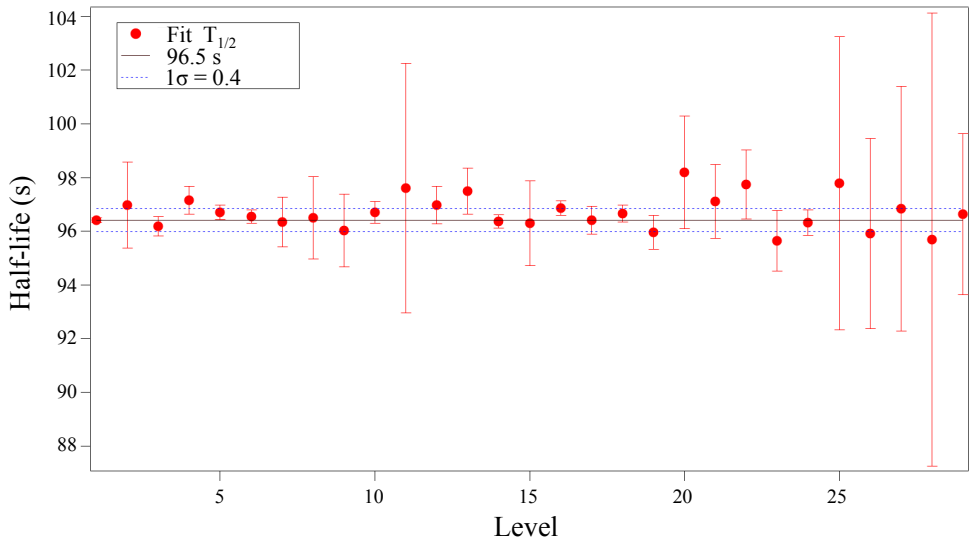}
\vspace*{-5cm}
\caption{\label{fig:otherGammas} The $T_{1/2}$ resulting from a fit for the strongest $\gamma$ ray depopulating each excited state observed in $^{46}$Ca (when statistically possible to fit). The data plotted here are also listed in Table~\ref{tab:otherGammas}. Comparisons of these values to the value of $T_{1/2}$=96.5 s, the plotted solid black line, show no indication that there is any contamination or isomeric states present. The dashed blue line represents the 1$\sigma$ error of 0.4 s from the weighted average of $T_{1/2}$ = 96.5 s considering all of the fit values. } 
\end{figure*}

\begin{table*}
\centering
\caption{\label{tab:otherGammas}The resultant $T_{1/2}$ for the most intense $\gamma$ ray depopulating select excited states observed in $^{46}$Ca. The listed numbers correspond to the plotted data shown in Fig.~\ref{fig:otherGammas}.}
\begin{tabular}{|c|ccc||c|ccc|}
\hline\hline
$\#$&E$_{Level}$ (keV) & $E_{\gamma}$ (keV)& $T_{1/2}$ (s) & $\#$& E$_{Level}$ (keV) & $E_{\gamma}$ (keV) &$T_{1/2}$ (s)\\
\hline
1 & 1346 & 1346 & 96.41(10) & 16 & 5375 & 4028 & 96.86(28)\\
2 & 2422 & 1076 & 97.0(16) & 17 & 5414 & 4068 & 96.4(53)\\
3 & 2575 & 1229 & 96.18(37) & 18 & 5535 & 4188 & 96.65(33)\\
4 & 3021 & 1675 & 97.15(51) & 19 & 5713 & 4366 & 95.95(64)\\
5 & 3611 & 2264 & 96.70(27) & 20 & 5815 & 4470 & 98.2(21)\\
6 & 3638 & 2292 & 96.54(25) & 21 & 5848 & 4501 & 97.1(14)\\
7 & 3858 & 2509 & 96.33(92) & 22 & 6032 & 4686 & 97.7(13)\\
8 & 3984 & 1409 & 96.5(15) & 23 & 6111 & 6110 & 95.6(11)\\
9 & 4386 & 776 & 96.0(14) & 24 & 6303 & 3727 & 96.33(48)\\
10 & 4404 & 3058 & 96.70(41) & 25 & 6378 & 5032 & 97.8(55)\\
11 & 4428 & 4428 & 97.6(47) & 26 & 6540 & 5190 & 95.9(36)\\
12 & 4432 & 1858 & 96.98(70) & 27 & 6705 & 5355 & 96.9(46)\\
13 & 4487 & 4486 & 97.50(85) & 28 & 6839 & 5492 & 95.7(84)\\
14 & 5052 & 3706 & 96.38(24) & 29 & 7034 & 5679 & 96.6(30)\\
15 & 5216 & 5216 & 96.3(16) & &&&\\
\hline

\end{tabular}
\end{table*}

To determine the $T_{1/2}$ of the $^{46}$K 2$^-$ ground state, the time-profiles of decays of $\gamma$ rays from $^{46}$Ca were investigated. The data analyzed contained an extended beam-off decay period of 1200 s that was binned into 1 s samples. The fits were performed starting at 300 s after the beginning of the decay period to minimize contributions from pile-up and high-rate effects that may have been present. Additionally, the behavior of the Compton-scattering background in the region of each $\gamma$ ray investigated was fit with an exponential and added to the overall fit as an energy-dependent background. An example of the fit of the decay of the 2$_1^+$ $\rightarrow$ 0$_1^+$ 1346 keV $\gamma$ ray is presented in Fig.~\ref{fig:halflife}.

The high statistics of these data made it possible to fit the decays of $\gamma$ rays depopulating 29 of the 42 excited states observed. The result of each fit is listed in Table~\ref{tab:otherGammas} and is plotted in Fig.~\ref{fig:otherGammas}. To look for the presence of systematic effects, the data were rebinned into 2  and 4 s bins and a ``chop analysis'' that changed the fit region was used.  This type of analysis is described in more detail in Ref. \cite{ryan}. From these investigations, a systematic uncertainty of 0.4 s is reported. The weighted average of all of the fits is $T_{1/2}$ = 96.5(4) s, which is in agreement with the precise value reported by Kunz $et~al.$

\begin{figure*}[t]
\centering
\hspace*{-0.5cm}
\includegraphics[scale=0.7]{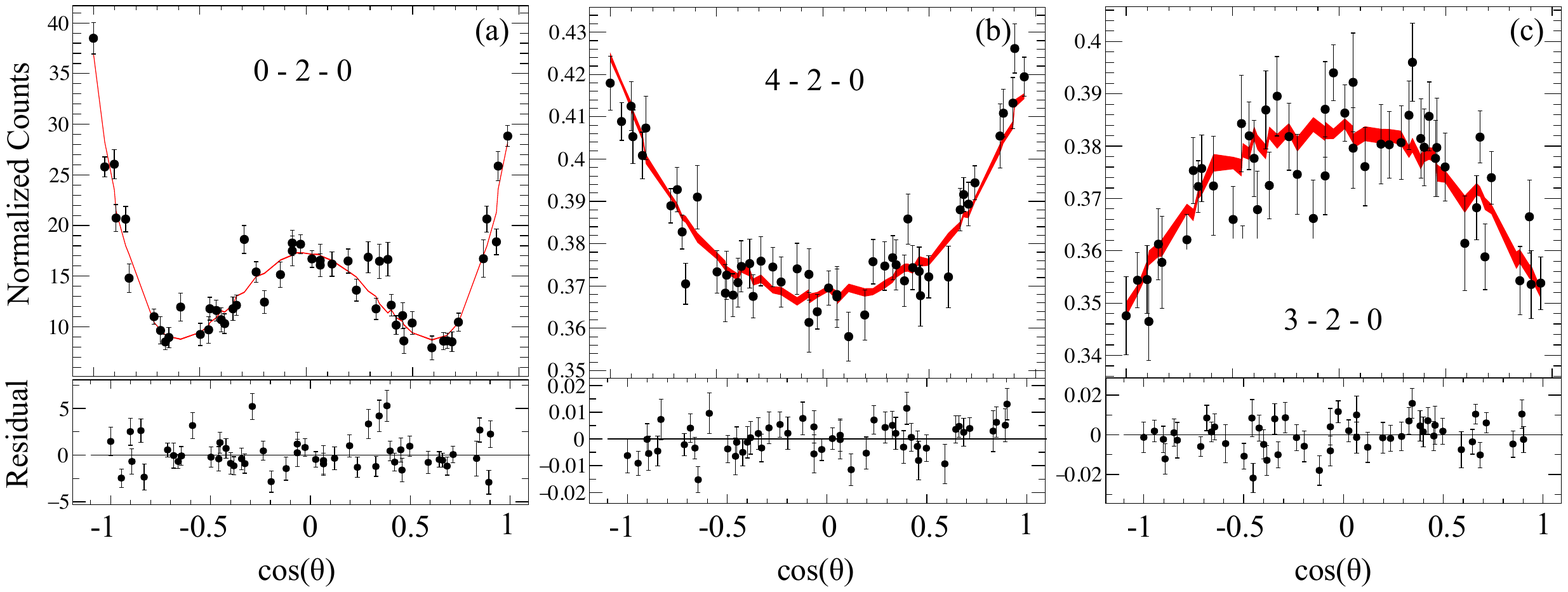}
\vspace*{-7cm}
\caption{\label{fig:2422cas} Three best-fit $\gamma$ - $\gamma$ angular correlations along with their respective residuals are presented to show the quality of the data. (a) The 0-2-0 correlation is for the 1076-1346 keV $\gamma$-ray cascade. (b) The 4-2-0 correlation is for the 1229-1346 keV $\gamma$-ray cascade. (c) The 3-2-0 correlation is for the 2264-1346 keV $\gamma$-ray cascade. For each angular correlation shown the width of the red line represents the statistical uncertainty of the fitted \textit{GEANT4} simulation and the statistical uncertainty of the angular correlation is shown as the black error bars for each point.  }
\end{figure*}

\section{$\gamma$-$\gamma$ Angular Correlations}
A directional correlation between two $\gamma$ rays, emitted in succession, is related  to the spins of the nuclear states as well as the multipolarities and multipole mixing ratios ($\delta$) of the transitions involved in the cascade. In general, the angular correlation between two successive $\gamma$ rays can be written as 
\begin{equation}\label{eq:LP}
W(\theta) = \sum^{\infty}_{i=0, even} B_{ii}G_{ii}(t)A_{ii}P_i(\cos\theta),
\end{equation}
where $B_{ii}$ indicate the initial nuclear orientation, $G_{ii}(t)$ are time-dependent perturbation factors that distort the angular correlation, $P_i(\cos\theta)$ are Legendre polynomials where $\theta$ is the angle between the $\gamma$ rays, and $A_{ii}$ are a series of coefficients \cite{angCorr}. In this scenario, the initial orientation of the nuclei is isotropic such that $B_{ii}$ = 1, and due to the short lifetimes of the excited states $G_{ii}$ $\approx$ 1. Then, taking into account the typically low spins and multipolarities involved in $\gamma$-ray cascades, Eqn.~\ref{eq:LP} can be simplified to 
\begin{equation}\label{eq:corr}
W(\theta) = A_{00}[1 + a_2P_2(\cos\theta) + a_4P_4(\cos\theta)],
\end{equation}
where 
\begin{equation}
a_i = A_{ii}/A_{00}
\end{equation}

An angular correlation analysis was performed on this data set to either assign or constrain spins for excited states and to determine $\delta$ values for the transitions. The 60 HPGe crystals were grouped into detector pairs reflecting the 51 unique angles of the GRIFFIN array ranging between 19$^{\circ}$ and 180$^{\circ}$.  When two successive $\gamma$ rays are detected in coincidence, the angle between them is simply the angle between the two crystals in which the $\gamma$ rays were detected. A detailed description of how physical angular correlations are extracted from what is observed experimentally with GRIFFIN is available in Ref.  \cite{jenna2}.

%%%%%%%%%%%%%%%%%%%%%%AngCorr Table%%%%%%%%%%%%%%%%%%%%%%%%%%
\begin{table*}[!ht]
\caption{\label{tab:angCorr}A summary of the $\gamma$-$\gamma$ angular correlation analysis. The spin assignments of selected excited states were investigated using the angular correlations of a cascade of two coincident $\gamma$ rays, $\gamma_1$  and $\gamma_2$. Twelve spin assignments ($J_{Lit.}$) from previous measurements are reported in the literature \cite{nndc}. The spin assignments ($J_{fit}$) from this work are shown in comparison; in some cases, $J_{fit}$ could not be assigned and the corresponding fit parameters for multiple values are shown. It was assumed that $\gamma_1$ was a transition of pure multipolarity with $\delta_1$ = 0, and that $\gamma_2$ could be a mixed transition described by $\delta_2$. The investigated spin for each $\gamma$-ray cascade is denoted with an asterisk(*). The resultant best-fit $a_2$, $a_4$, and $\delta_2$ parameters are reported with the corresponding $\chi^2$/$\nu$.Note that for these data all 4* $\rightarrow$ 2 $\rightarrow$ 0 transitions are assumed to have pure $E$2 multipolarity and are only listed if the fitted $\delta$ is 0.}
\begin{tabular}{|c|c|c|ccc|lccc|}
\hline
E$_{level}$ & $J_{Lit.}$ &$J_{fit}$& $\gamma_1$ (keV) & $\gamma_2$ (keV) &Cascade  & $\delta_2$& $a_2$ & $a_4$ & $\chi^2$/$\nu$\\
\hline
2422 & 0 &0& 1346 & 1076 & 0* $\rightarrow$ 2 $\rightarrow$ 0 & 0.00(2) & 0.481(33) & 1.22(5) & 1.41 \\
\hline
2575& 4 &4 &1346 & 1229 & 4* $\rightarrow$ 2 $\rightarrow$ 0& 0.00(2) &0.097(5) & 0.019(7) & 1.15\\
\hline
3021&2&2&1346&1675& 2* $\rightarrow$ 2 $\rightarrow$ 0& -0.17(5)&0.363(9)&0.023(13)&1.42 \\
\hline
3611&3&3&1346&2264& 3* $\rightarrow$ 2 $\rightarrow$ 0& 0.02(5)&-0.065(6)&-0.008(9)&1.30\\
\hline
3638&2&2&1346&2292& 2* $\rightarrow$ 2 $\rightarrow$ 0& 0.13(2)&0.151(14)&0.035(22)&1.17\\
\hline
3858&4&4&1346&2510&4* $\rightarrow$ 2 $\rightarrow$ 0&0.00(3)&0.121(11)&-0.057(17)&1.35\\
\hline
4257&&3&1346&2911&3* $\rightarrow$ 2 $\rightarrow$ 0& 0.38(4)&0.168(11)&-0.049(17)&1.51\\
&&3&1346&2911&3* $\rightarrow$ 2 $\rightarrow$ 0& 1.63(2)& --- & --- &1.40\\
\hline
4386&&2&1346&3040& 2* $\rightarrow$ 2 $\rightarrow$ 0&0.02(2)&0.240(6)&-0.028(9)&1.05\\
&&3&1346&3040& 3* $\rightarrow$ 2 $\rightarrow$ 0&0.61(3)&---&---&0.90\\
&&3&1346&3040& 3* $\rightarrow$ 2 $\rightarrow$ 0&1.13(3)&---&---&0.95\\
&&2&2264&776& 2*$\rightarrow$ 3 $\rightarrow$ 2&-0.24(1)&0.029(13)&0.012(19)&0.97\\
&&3&2264&776& 3*$\rightarrow$ 3 $\rightarrow$ 2&-3.25(3)&---&---&1.07\\
&&3&2264&776& 3*$\rightarrow$ 3 $\rightarrow$ 2&0.36(3)&---&---&0.95\\
\hline
4404&3&3&1229&1830& 3* $\rightarrow$ 4 $\rightarrow$ 2&-0.02(2)&-0.125(24)&0.054(35)&0.84\\
&&4&1229&1830& 4* $\rightarrow$ 4 $\rightarrow$ 2&1.03(6)&---&---&0.85\\
\hline
4428&2&1&4428&624& 2 $\rightarrow$  1* $\rightarrow$ 0&-0.44(5)&0.321(44)&-0.064(63)&1.45\\
&&2&4428&624& 2 $\rightarrow$  2* $\rightarrow$ 0&-0.13(9)&---&----&1.45\\
&&3&4428&624& 2 $\rightarrow$  3* $\rightarrow$ 0&-0.41(3)&---&---&1.47\\
\hline
4432&2&3&1229&1858&3* $\rightarrow$ 4 $\rightarrow$ 2&-0.03(2)&-0.124(23)&-0.038(34)&0.92\\
&&4&1229&1858&4* $\rightarrow$ 4 $\rightarrow$ 2&1.00(6)&---&---&1.17\\
\hline
4487&(4)&2&1346&3142&2* $\rightarrow$ 2 $\rightarrow$ 0&0.15(1)&0.139(12)&0.016(17)&1.07\\
&&3&1346&3142&3* $\rightarrow$ 2 $\rightarrow$ 0&0.31(3)&---&---&1.11\\
&&4&1346&3142&4* $\rightarrow$ 2 $\rightarrow$ 0&0.00(6)&---&---&1.08\\

\hline
5052&(4)&2&1346&3706& 2* $\rightarrow$ 2 $\rightarrow$ 0&-0.02(2)&0.266(3)&-0.001(5)&1.50\\
\hline
5375&(3)&3&1346&4029&3* $\rightarrow$ 2 $\rightarrow$ 0&0.00(2)&-0.075(6)&0.013(9)&1.28\\
&&3&1229&2800&3* $\rightarrow$ 4 $\rightarrow$ 2&0.00(2)&---&---&0.69\\
&&4&1229&2800&4* $\rightarrow$ 4 $\rightarrow$ 2&1.00(9)&---&---&0.81\\
\hline
5414&&1&2264&1804&1* $\rightarrow$ 3 $\rightarrow$ 2&0.32(6)&0.192(31)&-0.012(45)&1.09\\
&&1&2264&1804&1* $\rightarrow$ 3 $\rightarrow$ 2&1.96(7)&---&---&1.03\\
&&2&2264&1804&2* $\rightarrow$ 3 $\rightarrow$ 2&-0.31(2)&---&---&1.14\\
&&3&2264&1804&3* $\rightarrow$ 3 $\rightarrow$ 2&-2.13(3)&---&---&0.98\\
&&3&2264&1804&3* $\rightarrow$ 3 $\rightarrow$ 2&0.21(3)&---&---&1.10\\
\hline
5535&(4)&2&1346&4189&2* $\rightarrow$ 2 $\rightarrow$ 0&0.38(2)&-0.044(10)&-0.004(16)&1.30\\
&&3&1346&4189&3* $\rightarrow$ 2 $\rightarrow$ 0&0.03(1)&---&---&1.10\\

\hline
5712&&2&1346&4366&2* $\rightarrow$ 2 $\rightarrow$ 0&0.06(2)&0.200(19)&-0.011(28)&1.43\\
&&3&1346&4366&3* $\rightarrow$ 2 $\rightarrow$ 0&0.45(5)&---&---&1.43\\
&&3&1346&4366&3* $\rightarrow$ 2 $\rightarrow$ 0&1.47(5)&---&---&1.49\\

\hline
6111&&2&1346&4765&2* $\rightarrow$ 2 $\rightarrow$ 0&0.70(3)&-0.216(27)&0.056(40)&1.42\\
&&3&1346&4765&3* $\rightarrow$ 2 $\rightarrow$ 0&-0.17(5)&---&---&1.42\\

\hline
\end{tabular}
\end{table*}
%%%%%%%%%%%%%%%%%%%%%%%%%%%%%%%%%%%%%%%%%%%

For the angular correlations examined, two of the three spins ($J$) of the involved nuclear states were previously reported, and one of the two $\gamma$ rays had pure multipolarity ($\delta$ = 0). This scenario allowed for the unknown spin of the third nuclear state to be assigned or constrained and for the $\delta$ value of the second $\gamma$ ray to be determined. Analysis of each angular correlation was performed using Method 2 from \cite{jenna2}, where the experimental data are fitted with a \textit{GEANT4} simulation that corrects for the finite detector sizes of the GRIFFIN HPGe detectors and also takes into account many different combinations of potential spins and mixing ratios for a given pair of $\gamma$-ray energies. For each angular correlation the values of $a_2$ and $a_4$ were fitted directly as they are spin independent. To determine mixing ratios, the values of the unknown spin were varied from 0 to 5 and the best fits of $\delta$ for the second $\gamma$ ray for each spin were determined given a $\chi^2$ minimization. For a $J$ value to be considered for assignment, the fitted minimum $\chi^2$ must be within the 99\% confidence interval, given the number of degrees of freedom, for the fit. In several cases, the fitted values of $\delta$ for multiple $J$ values satisfied this constraint. Results of this analysis are shown in Table~\ref{tab:angCorr}. To highlight the quality of the data and the sensitivity of this technique,  angular correlations are presented in Fig.~\ref{fig:2422cas} for $J_i$-$J_x$-$J_f$= 0-2-0, 4-2-0, and 3-2-0 $\gamma$-ray cascades where roughly 10$^5$, 1.5$\times$10$^6$, and 10$^6$  $\gamma$-$\gamma$ coincidences were observed, respectively. 

\section{Discussion of $J^{\pi}$ Assignments}\label{jpi}
In this study, 42 excited states were placed in the level scheme of $^{46}$Ca. Previous spin assignments had been made for 19 of these states from various transfer reaction measurements \cite{nndc}. From the analysis of these data, information on the spin and parity ($J^{\pi}$) of the excited states can be deduced from a combination of the observed $\beta$-feeding log$ft$ values with $\gamma$-$\gamma$ angular correlations and decay systematics within the level scheme. Assigned or constrained $J^{\pi}$ values are listed in Table~\ref{tab:bigTable}. 

In this measurement, the excited states in $^{46}$Ca were populated via the $\beta^-$ decay of the $J^{\pi}$ =  2$^-$ $^{46}$K ground state. From $\beta$-decay selection rules, the log$ft$ value determined for a particular decay reflects the change in angular momentum between the initial decaying state of the parent and the final populated state of the daughter nucleus, as well as whether or not there was a change in parity between the two states.  As 40 of the 42 observed excited states in $^{46}$Ca were populated directly via $\beta$ feeding, it was possible to make constrained $J^{\pi}$ assignments for these states given the observed amount of feeding to each state. Of course, it is difficult to make constraints on $J^{\pi}$ solely given the observed $\beta$ feeding, but for the purposes of this work, constraints on $J^{\pi}$ are made given lower limits to log$ft$ values that correspond to allowed, first-forbidden, unique first-forbidden, and second-forbidden $\beta$ transitions of log$ft$ > 4, 5, 8, and 11, respectively. If the spin of an excited state could be investigated via $\gamma$-$\gamma$ angular correlations, further constraints or a singular assignment was made. Also, in some cases further constraints were made based on decay systematics within the level scheme or the likeliness of a fitted $\delta$ value for a given type of transition . Discussion of assignments and constraints for select excited states follows.

\subsection{The 4257 keV Level}
The 4257 keV excited state is reported for the first time in this work. The angular correlation of the 2911-1346 keV $\gamma$-ray cascade, depopulating the 4257 keV state to the $J^{\pi}$ = 0$^+$ ground state, was investigated. As is reported in Table~\ref{tab:angCorr}, a $\chi^2$ minimization is observed for a $J$ = 3 assignment. As a result of this fits in conjunction with the observed $\beta$ feeding (log$ft$ = 8.20(5)) to this state,  a $J^{\pi}$ assignment of 3$^+$ is made in this work.

\subsection{The 4386 keV Level}
\begin{figure}
\vspace{-3cm}
\hspace*{-1cm}
\includegraphics[scale=0.77]{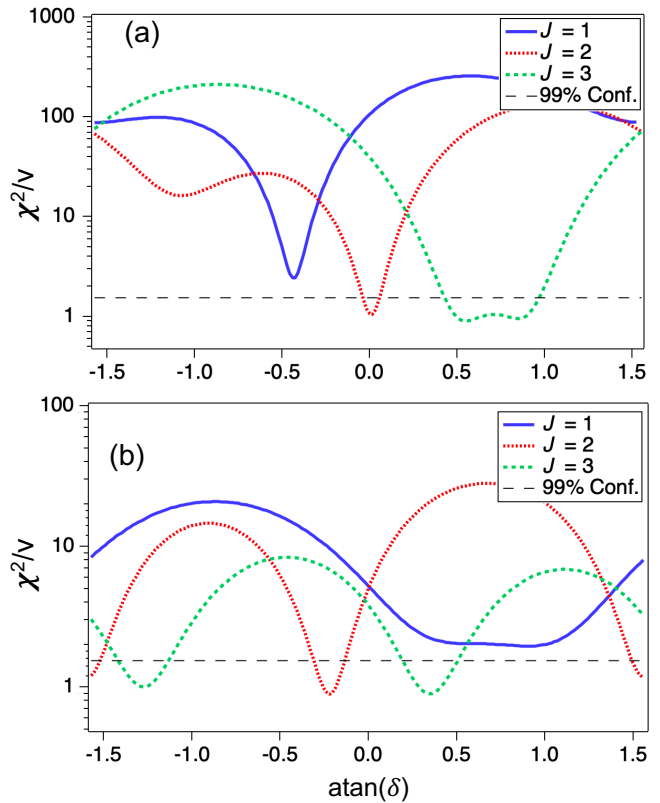} 
\vspace{-3.9cm}
\caption{\label{fig:4386level} The 3040-1346 and 776-2264 keV $\gamma$-ray cascades both depopulate the 4386 keV excited state in $^{46}$Ca. The two plots here show the comparison of $\chi^2$/$\nu$ for possible values of $J$ = 1, 2, and 3 with all possible values of $\delta$ along with the 99\% confidence interval for each cascade. (a) Fits for the 3040-1346 keV cascade. The best fit occurs for the  $J$ = 2 and 3 assignments. (b) Fits for the 776-2264 keV cascade. Best fits for the $J$ = 2 and 3 assignments both fall within the 99\% confidence interval.  }
\end{figure}

The 4386 keV excited state is reported for the first time in this work. The angular correlations of two $\gamma$-ray cascades were investigated, the first being the 3040-1346 keV cascade (Fig.~\ref{fig:4386level}a.) and the second being the 776-2264 keV cascade  (Fig.~\ref{fig:4386level}b.). Plots of $\chi^2$/$\nu$ for possible values of $J$ with all possible values of $\delta$ are shown. As is reported in Table~\ref{tab:angCorr}, $\chi^2$ minimizations were observed for $J$ = 2 and 3 assignments. The result of these fits in conjunction with the observed $\beta$ feeding (log$ft$ = 6.44(2)) to this state show that $J^{\pi}$ assignments of (2, 3)$^+$ and (2, 3)$^-$  are possible. However, as the 4386 keV level populates the ground state,  potential $J^{\pi}$ assignments of 2$^-$ and 3$^+$ can be ruled out. Additionally, a 3$^-$ assignment is not possible as the fitted value of $\delta$ = 0.61(3) is much too large for a 3$^-$ $\rightarrow$ ~2$^+$  transition that should be pure $E$1.  For the 4386 keV, level a $J^{\pi}$ assignment of 2$^+$ is made.

 %%%%%%%%%%%%%%%%%%%
 %%%%%%%%%%%%%%%%%%%%

\subsection{The 4404 keV Level}
Previously, a states at 4406.0(15) and 4409(3) keV were reported from ($p,p')$ and ($p,t$) reaction studies \cite{pp,pt} and a $J^{\pi}$ = 3$^-$ assignment was made \cite{pt}. The angular correlation of the 1830-1229 keV $\gamma$-ray cascade, which depopulates the 4404 keV state to the $J^{\pi}$ = 2$^+$ 1346 keV state, via the $J^{\pi}$ = 4$^+$ 2575 keV state, was investigated. As is reported in Table~\ref{tab:angCorr}, $\chi^2$ minimizations were observed for $J$ = 3 and 4 assignments. However, the potential $J$ = 4 assignment can be ruled out as the 4404 keV level directly populates the $J^{\pi}$ = 0$^+$ ground state. Given the observed $\beta$ feeding (log$ft$ = 6.37(2)) to this state, an assignment of $J^{\pi}$ = 3$^-$ is reported for the 4404 keV level. Note that the fitted value of $\delta$ = -0.02(4) is consistent with a 3$^-$ $\rightarrow$ ~4$^+$ transition that should have pure $E$1 multipolarity.

\subsection{The 4428 keV Level}
A $J^{\pi}$ = 2 spin assignment for a level at 4433(7) was observed in a previous ($t,p$) study  \cite{tp}.  An angular correlation analysis was performed on the 624-4428 keV $\gamma$-ray cascade, which depopulates the 4428 keV state to the $J^{\pi}$ = 0$^+$ ground state. As is reported in Table~\ref{tab:angCorr}, $\chi^2$ minimizations were observed for $J$ = 1, 2, 3.  The result of these fits in conjunction with the observed $\beta$ feeding (log$ft$ = 8.80(4)) to this state show that $J^{\pi}$ assignments of (1, 2, 3)$^+$ are possible. However, as this state does directly populate the $J^{\pi}$ = 0$^+$ ground state,  potential $J^{\pi}$ = 3$^+$  assignment can be eliminated. Additionally, the 1$^+$ assignment can be ruled out as the fitted value of $\delta$ = -0.44(5) is much too large for a 2$^-$ $\rightarrow$ ~1$^+$ transition that should be pure $E$1. For the 4428 keV level, a $J^{\pi}$ assignment of 2$^+$ is, therefore, concluded.

\subsection{The 4432 keV Level}
There has been no previous observation of an excited state at 4432 keV. The angular correlation of the 1868-1229 keV $\gamma$-ray cascade, which depopulates the 4432 keV state to the $J^{\pi}$ = 2$^+$ 1346 keV state, via the $J^{\pi}$ = 4$^+$ 2575 keV state, was investigated. As is reported in Table~\ref{tab:angCorr}, $\chi^2$ minimizations were observed for $J$ = 3 and 4. From the $\beta$-feeding intensity to this state of 2.59(7)\% log$ft$ values of 6.65(2) and 8.12(2) can both be calculated given that the transition is or is not unique, respectively, such that $J^{\pi}$ = 3$^+$, 3$^-$, and 4$^+$ assignments are possible. For the 4432 keV level, a constrained $J^{\pi}$ assignment of (3, 4$^+$) is made.

\subsection{The 4487 keV Level}
Previously, a state at 4493(3) keV was reported from a ($p,t$) transfer reaction and a tentative $J^{\pi}$ = (4$^+$) assignment was made \cite{pt}. The angular correlation of the 3142-1346 keV $\gamma$-ray cascade, which populates the $J^{\pi}$ = 0$^+$ ground state, via the $J^{\pi}$ = 2$^+$ 1346 keV state, was investigated. As is reported in Table~\ref{tab:angCorr}, $\chi^2$ minimizations were observed for $J$ = 2, 3, and 4.  The result of the fits of the angular correlation in conjunction with the observed $\beta$ feeding (log$ft$ = 7.36(2)) to this state show that $J^{\pi}$ assignments of (2, 3)$^+$ and (2, 3)$^-$  are possible. The 4487 keV level populates the ground state, thus potential $J^{\pi}$ assignments of 2$^-$ and 3$^+$ can be ruled out. Additionally, the 3$^+$ assignment can be eliminated as the fitted value of $\delta$ = 0.31(3) is much too large for a 3$^-$ $\rightarrow$ ~2$^+$ $\gamma$-ray transition that should be pure $E$1. For the 4487 keV level, a $J^{\pi}$ assignment of 2$^+$ is concluded. 

\begin{figure}
\vspace{-2.8cm}
\hspace*{-1cm}
\includegraphics[scale=0.8]{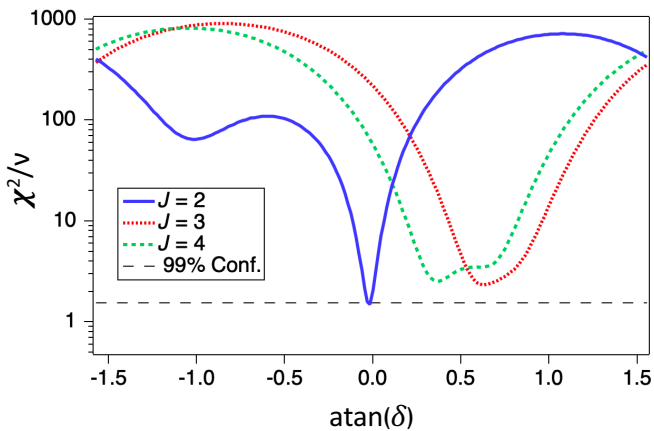} 
\vspace{-9.4cm}
\caption{\label{fig:5052level} The 3706-1346 $\gamma$-ray cascade depopulates the 5052 keV excited state in $^{46}$Ca. The plot shows the comparison of $\chi^2$/$\nu$ for possible values of $J$ =  2, 3, 4, and 5 with all possible values of $\delta$ along with the 99\% confidence interval. The best fit occurs for the  $J$ = 2 assignment. }
\end{figure}

\subsection{The 5052 keV Level}
A tentative spin of  $J^{\pi}$ = (4$^+$) has been adopted \cite{pt} for a state at 5052 keV in a $^{48}$Ca($p,t$) measurment. In a previous $\beta$-decay measurement of the $^{46}$K 2$^-$ ground state, an excited state at 5047 keV (no uncertainty reported) was given a tentative spin assignment of $J^{\pi}$ = (2$^-$) or (3$^-$) from observed log$ft$ values \cite{parsa}. In the present work, the 5052 keV state was found to be the most intensely populated state from the $\beta^-$ decay of the $^{46}$K 2$^-$ ground state ($I_{\beta}$ = 39.8(13)$\%$). From the observed log$ft$ = 5.06(2), spin assignments of $J^{\pi}$ = 1$^-$, 2$^-$ or 3$^-$ are the most probable. The angular correlation of the 3706-1346 keV $\gamma$-ray cascade, depopulating the 5052 keV state, was investigated and is shown in Fig.~\ref{fig:5052level} . The minimum fitted $\chi^2$ was observed  for a $J$ = 2 assignment with $\delta_{3706}$ = -0.02(2), consistent with a pure $E$1 transition, and therefore a $J^{\pi}$ = 2$^-$ assignment is made.

\subsection{The 5351 keV Level}
There has been no previous observation of a state at 5351 keV. This state is very weakly populated from the $\beta$ decay of the $J^{\pi}$ = 2$^-$  $^{46}$K ground state with a feeding intensity of less than  0.005\% observed in this measurement. Given the observed feeding intensity, lower limits on log$ft$ values of >8.73 and >9.96 can both be calculated given that the transition is or is not unique, respectively. From these, a potential assignment of  $J^{\pi}$ = (4$^+$) seems to be the most probable for the 5351 keV state.

\subsection{The 5375 keV Level}
Previously, a state at 5380(4) keV was reported from a ($p,t$) transfer reaction and a tentative $J^{\pi}$ = (3$^-$ ) assignment was made \cite{pt}. Angular correlations of the 4029-1346 keV and 2800-1229 keV $\gamma$-ray cascades, depopulating the 5375 state to the $J^{\pi}$ = 2$^+$ 1346 keV state and the $J^{\pi}$ = 4$^+$ 2575 keV state, respectively, were both analyzed. As is reported in Table~\ref{tab:angCorr}, $\chi^2$ minimizations were observed for $J$ = 3 and 4. As this state was observed to directly populate the $J^{\pi}$ = 0$^+$ ground state,  potential $J$ = 3$^+$ and 4$^+$ assignments can be eliminated. Given the observed $\beta$ feeding (log$ft$ = 6.68(2)) to this state, an assignment of $J^{\pi}$ = 3$^-$ is made for the 5375 keV level. 

\subsection{The 5414 keV Level}
There has been no previous observation of an excited state at 5414 keV. An angular correlation analysis was performed on the 1804-2264 keV $\gamma$-ray cascade, which populates the $J^{\pi}$ = 2$^+$ 1346 keV state via the $J^{\pi}$ = 3$^-$ 3611 keV state. As is reported in Table~\ref{tab:angCorr}, $\chi^2$ minimizations were observed for $J$ = 1, 2, and 3. The result of the fits of the angular correlation in conjunction with the observed $\beta$ feeding (log$ft$ = 5.91(2)) to this state show that $J^{\pi}$ assignments of (1, 2, 3)$^-$ are possible. However, as the 5414 keV state directly populates the $J^{\pi}$ = 0$^+$ ground state, the potential $J^{\pi}$ = 2$^-$ assignment can be ruled out. For the 5414 keV level, a constrained $J^{\pi}$ assignment of (1, 3)$^-$ is suggested. 

\subsection{The 5535 keV Level}
The 5538(4) keV state was observed in the $^{48}$Ca($p,t$) reaction \cite{pt}. The angular correlation of the 4189-1346 keV $\gamma$-ray cascade, which depopulates the 5535 keV state to the $J^{\pi}$ = 0$^+$ ground state via the $J^{\pi}$ = 2$^+$ 1346 keV state, was investigated. As is reported in Table~\ref{tab:angCorr}, $\chi^2$ minimizations were observed for $J$ = 2 and 3 assignments.  The result of the fits of the angular correlation in conjunction with the observed $\beta$ feeding (log$ft$ = 6.09(2)) to this state reveal that $J^{\pi}$ assignments of (2, 3)$^+$ and (2, 3)$^-$  are possible. The 5535 keV level populates the ground state, so potential $J^{\pi}$ assignments of 2$^-$ and 3$^+$ can be ruled out. For the 5535 keV level, a constrained $J^{\pi}$ assignment of (2$^+$, 3$^-$) is made. 

\subsection{The 5712 keV Level}
There has been no previous observation of an excited state at 5712 keV. The angular correlation of the 4366-1346 keV $\gamma$-ray cascade, which populates the $J^{\pi}$ = 0$^+$ ground state via the $J^{\pi}$ = 2$^+$ 1346 keV state, was investigated. As is reported in Table~\ref{tab:angCorr}, $\chi^2$ minimizations were observed for $J$ = 2 and 3. From the $\beta$-feeding intensity to this state of 2.59(7)\% and a log$ft$ value of 6.09(2), $J^{\pi}$ assignments of (2, 3)$^+$ and (2, 3)$^-$  are possible. A 3$^-$ assignment can be ruled out as fitted values of $\delta$ = 0.45(5) and 1.47(5) are  too large for a 3$^-$ $\rightarrow$ ~2$^+$  transition that should be pure $E$1.  For the 5712 keV level, a constrained $J^{\pi}$ assignment of (2, 3$^+$) is suggested. 

\subsection{The 6111 keV Level}
There has been no previous observation of an excited state at 6111 keV. An angular correlation analysis was performed on the 4765-1346 keV $\gamma$-ray cascade, which populates the $J^{\pi}$ = 0$^+$ ground state. As is reported in Table~\ref{tab:angCorr}, $\chi^2$ minimizations were observed for $J$ = 2 and 3. The result of the fits of the angular correlation in conjunction with the observed $\beta$ feeding (log$ft$ = 5.86(2)) to this state show that $J^{\pi}$ assignments of (2, 3)$^+$ and (2, 3)$^-$  are possible. As this state does directly populate the $J^{\pi}$ = 0$^+$ ground state, potential $J^{\pi}$ = 2$^-$  and 3$^+$ assignments can be eliminated. For the 6111 keV level, a constrained $J^{\pi}$ assignment of (2$^+$, 3$^-$) is concluded. 
 
\section{Theoretical Calculations and Interpretation of Results}
\begin{figure}
\vspace*{-3.cm}
\hspace*{-1.1cm}
\includegraphics[scale = 0.75]{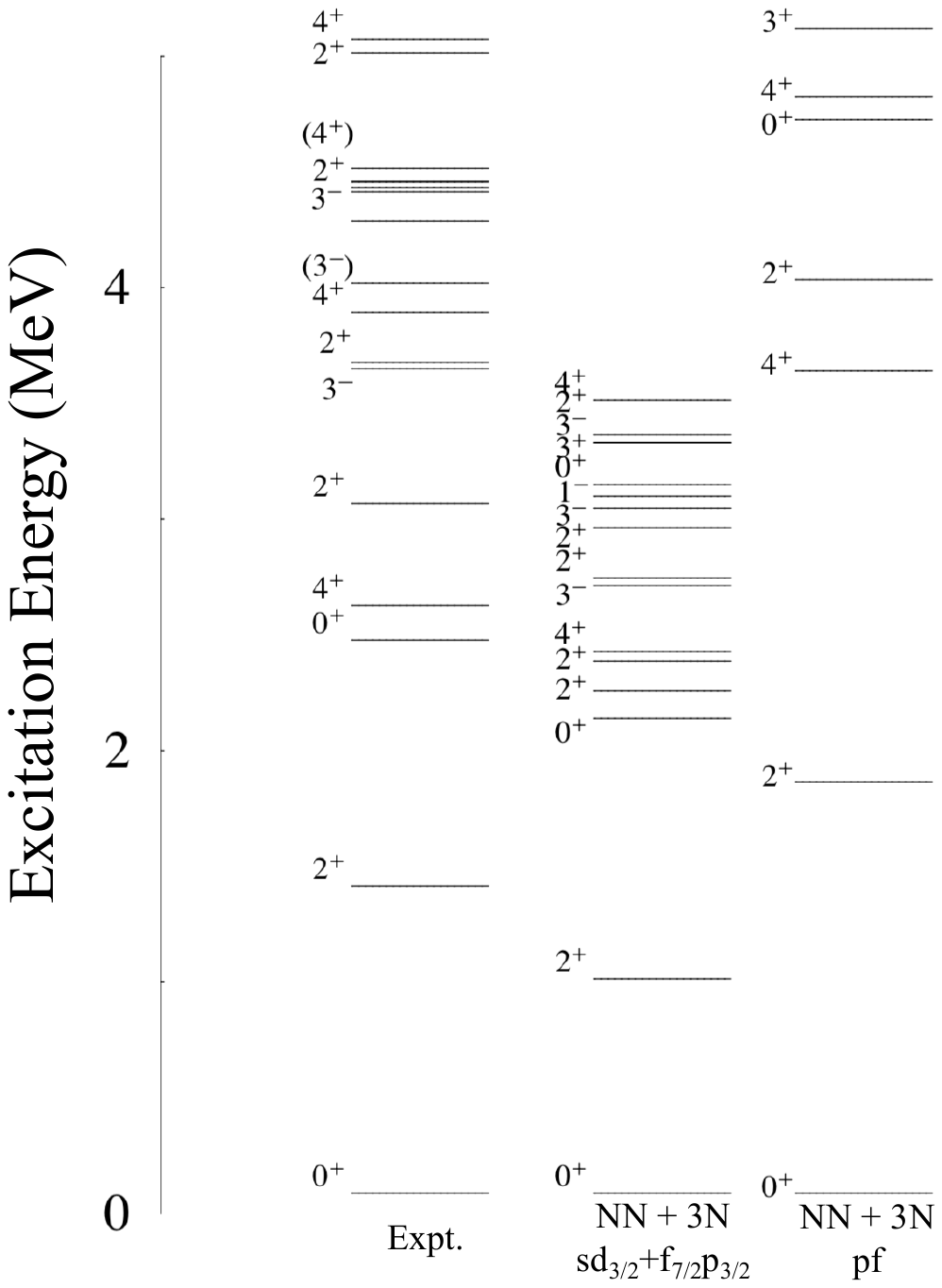}
\vspace*{-3cm}
\caption{\label{fig:calcs}The experimentally observed structure of $^{46}$Ca compared to theoretical predictions from two different microscopic valence-space Hamiltonians derived from NN+3N forces.}
\end{figure}
The level structure of $^{46}$Ca is compared to predictions from a microscopic valence-space Hamiltonian derived from the two-nucleon (NN) and three-nucleon (3N) forces of Refs. \cite{simo16, simo17, hebe11} and the perturbative many-body approach described in Ref.~\cite{jason2}. The NN interactions from chiral effective field theory \cite{chiralEFT, renorm} are evolved to low-momentum scales via similarity renormalization group methods \cite{bogner1}. Three-nucleon forces, fit to reproduce properties of few-body systems, are captured via normal ordering with respect to the relevant core nucleus. Excitations outside of the valence space are taken into account to third order in many-body perturbation theory within a harmonic-oscillator basis of 13 major shells. This approach has been shown to reproduce well the ground-state energies and spectroscopy in neutron-rich calcium isotopes \cite{jason1, Hebeler, jason2, weinholtz, jason3} as well as non-intruder states in lighter isotopes \cite{jason2}. To investigate the role of cross-shell configurations on the spectroscopy of $^{46}$Ca, we extend this approach to the non-standard valence space comprised of the proton $s_{1/2}$ $d_{3/2}$ $f_{7/2}$ $p_{3/2}$ and neutron $s_{1/2}$ $d_{3/2}$ $f_{7/2}$ $p_{3/2}$ $p_{1/2}$ single-particle orbitals above a $^{28}$Si core. Finally, we diagonalize the resulting valence-space Hamiltonian with the NuShellX shell-model code \cite{brown}. Previous calculations have been published for $^{46}$Ca \cite{jason2}, but these could not reproduce the existence of the first excited 0$^+$ state or obviously the negative-parity states since the proton $s_{1/2}$ $d_{3/2}$ single-particle orbitals were not included in the valence space.

The calculated levels are shown in comparison to the experimentally observed excited states in $^{46}$Ca in Fig.~\ref{fig:calcs}. In the $pf$-shell calculation, the spectrum is similar to the results discussed in Ref.~\cite{jason2}, where the first excited $2^+$ state is systematically several hundred keV high across the chain. Furthermore, the spectrum is much too spread compared to experiment, indicating the likely need for cross-shell degrees of freedom. When appropriate cross-shell orbitals are included in the valence space, however, the calculation is found to reproduce the existence of the negative-parity states and the 0$^+$ and $2^+$ intruder state. This result highlights that the configurations of these states are dominated by $sd$ proton excitations. The calculations also illustrate that the large cluster of nearly degenerate excited states present below 4~MeV are mainly due to cross-shell excitations, which reveals that protons are not as inert as might be expected for a semi-magic isotopic chain and clearly underpins the need to include such cross-shell degrees of freedom explicitly in the valence space. However it is clear that the calculated excited-state energies are systematically too compressed. This is likely due to the procedure for capturing 3N forces in the calculation. Since the core is now taken to be $^{28}$Si, and 3N forces are normal ordered with respect to the core only, we now neglect 3N forces between the 6 protons and 12 neutrons in the valence space. Including these repulsive effects would result in a more appropriately spaced spectrum.

Similar pictures have been drawn when analyzing electromagnetic moments and transitions in other systems with valence particles of only neutrons, such as the light calcium \cite{ruiz15} and nickel isotopes \cite{kenn00, evitts18}. While this has been well studied with phenomenological shell-model calculations, it is still a challenge for $ab~initio$ methods using the valence-space paradigm. For example, in this work, we performed calculations within the $ab~initio$ valence-space in-medium similarity renormalization group (VS-IMSRG) approach \cite{tsuk, bogner2, stroberg, stroberg2} for the same valence space, but the neutron $sd$-$pf$ shell gap was found to be much too large to give meaningful results. Work is currently in progress to better understand the decoupling of cross-shell valence spaces in this region and ultimately produce $ab~initio$ multi-shell interaction based only on NN+3N forces. The insights drawn from these calculations highlight the necessity of detailed experimental data, such as is provided by the present work, with which to benchmark and compare theoretical results and drive forward the ongoing development efforts.

\section{Summary}
The $^{46}$Ca level scheme has been greatly expanded as the result of a high-statistics measurement of the $\beta^-$ decay of the $^{46}$K $J^{\pi}$ = 2$^-$ ground state conducted with the GRIFFIN spectrometer at TRIUMF-ISAC. A more comprehensive level scheme was constructed from a $\gamma$-$\gamma$ coincidence analysis. An angular correlation analysis was utilized to confirm and assign spins of observed excited states. The results of theoretical calculations that utilize a microscopic valence-space Hamiltonian derived from two-nucleon and three-nucleon forces  reveal that the proton excitations play an important role in the structure of this nucleus. Additional effort is required to produce realistic $ab~initio$ predictions for nuclei in this region. 

\begin{acknowledgments}
We would like to thank the operations and beam delivery staff at TRIUMF for providing the radioactive beam. The GRIFFIN spectrometer was jointly funded by the Canadian Foundation for Innovation (CFI), TRIUMF, and the University of Guelph. TRIUMF receives federal funding via a contribution agreement through the National Research Council Canada (NRC). C.E.S. acknowledges support from the Canada Research Chairs program. This work was supported in part by the Natural Sciences and Engineering Research Council of Canada (NSERC). This material is based upon work supported by the U.S. National Science Foundation under Grant No. PHY-1913028.
\end{acknowledgments}

\begin{figure*}
\centering
    
		\includegraphics[scale=0.8]{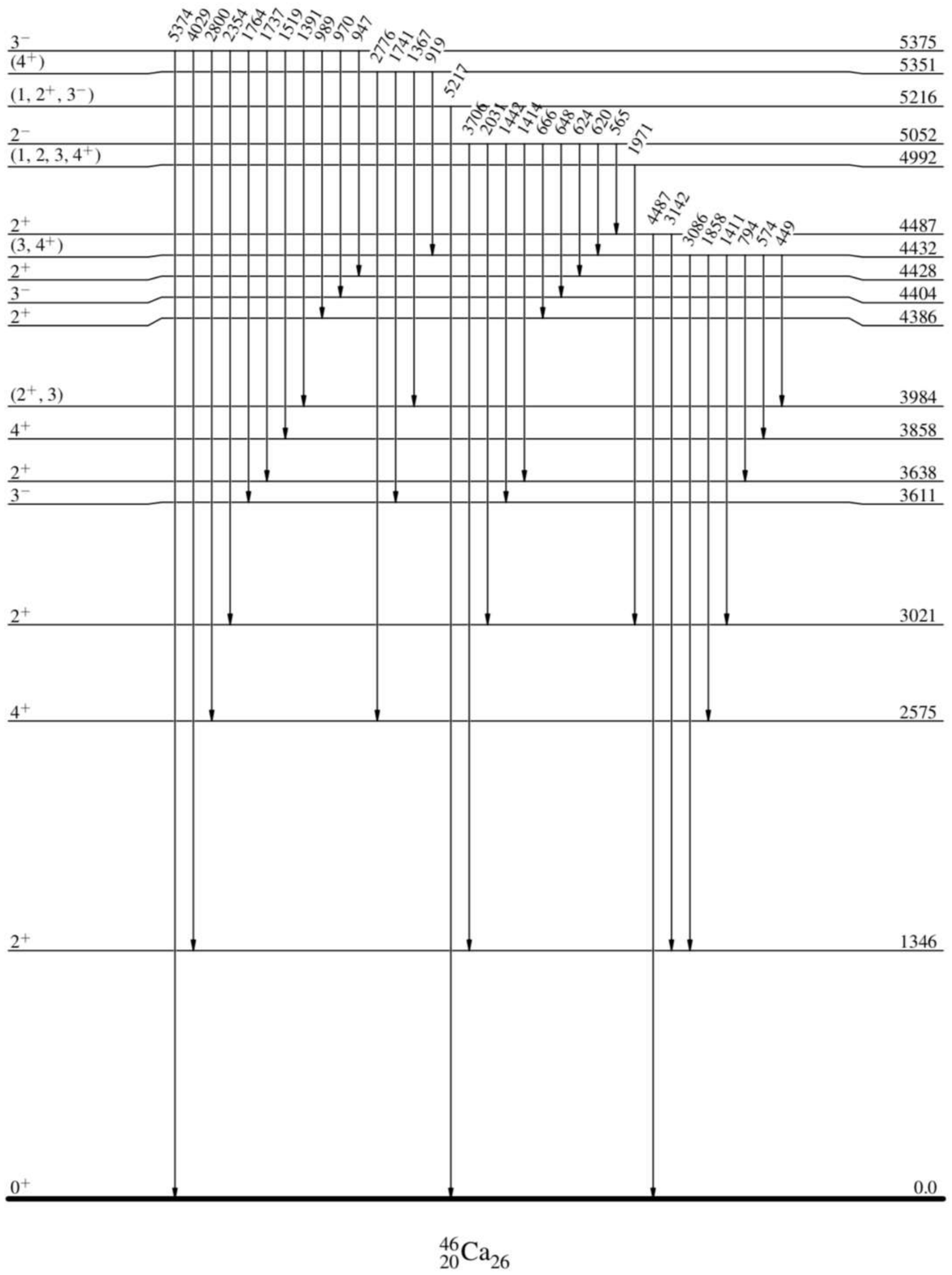}
		\caption{A partial level scheme of levels populated in $^{46}$Ca from the $\beta^-$ decay of the $J^{\pi}$ = 2$^- $ $^{46}$K ground state showing $\gamma$ rays that depopulate the 4432-5375 keV levels.}
	
\end{figure*}

\begin{figure*}
\centering
    
		\includegraphics[scale=0.8]{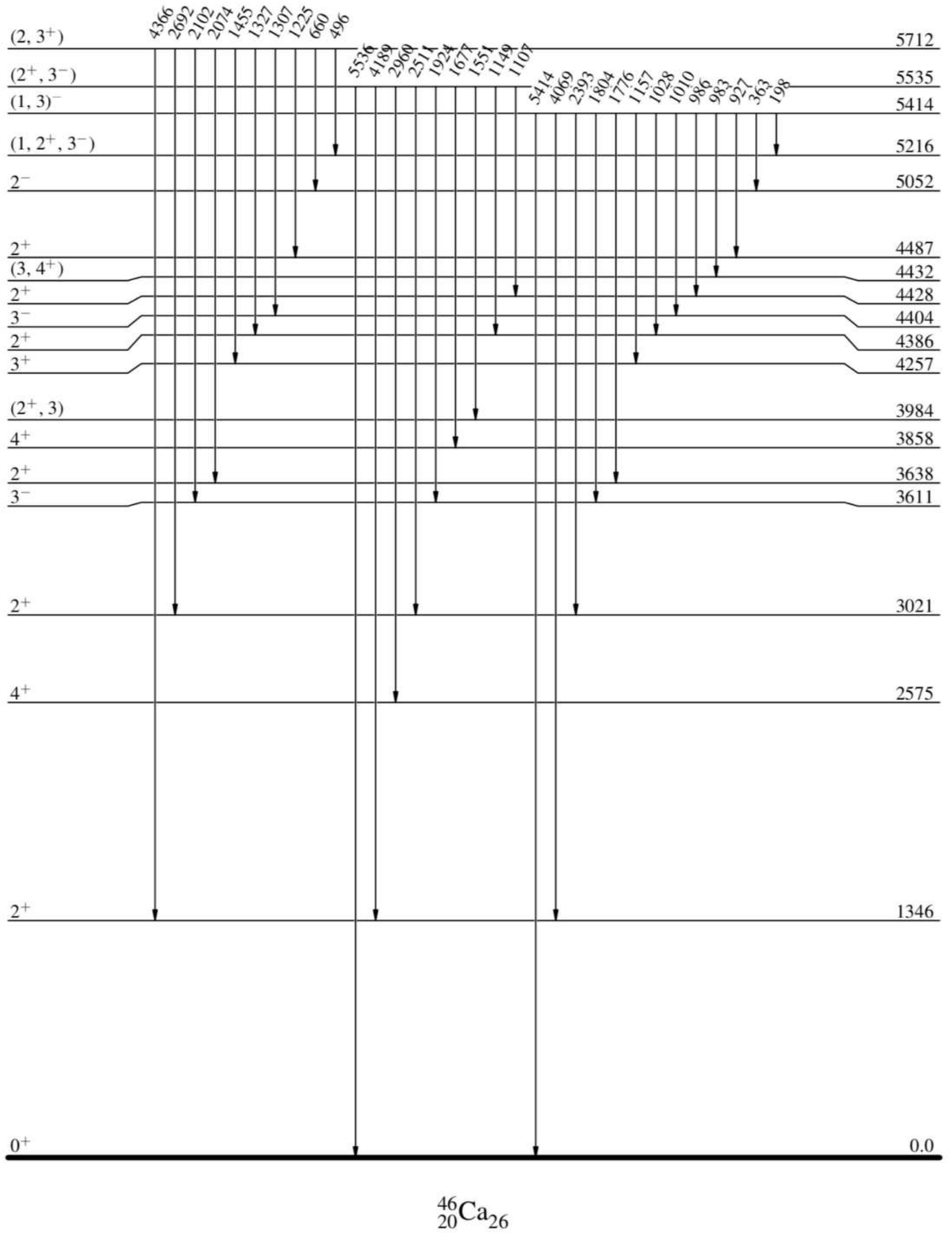}
		\caption{A partial level scheme of levels populated in $^{46}$Ca from the $\beta^-$ decay of the $J^{\pi}$ = 2$^- $ $^{46}$K ground state showing $\gamma$ rays that depopulate the 5414-5712 keV levels.}
	
\end{figure*}

\begin{figure*}
\centering
    
		\includegraphics[scale=0.8]{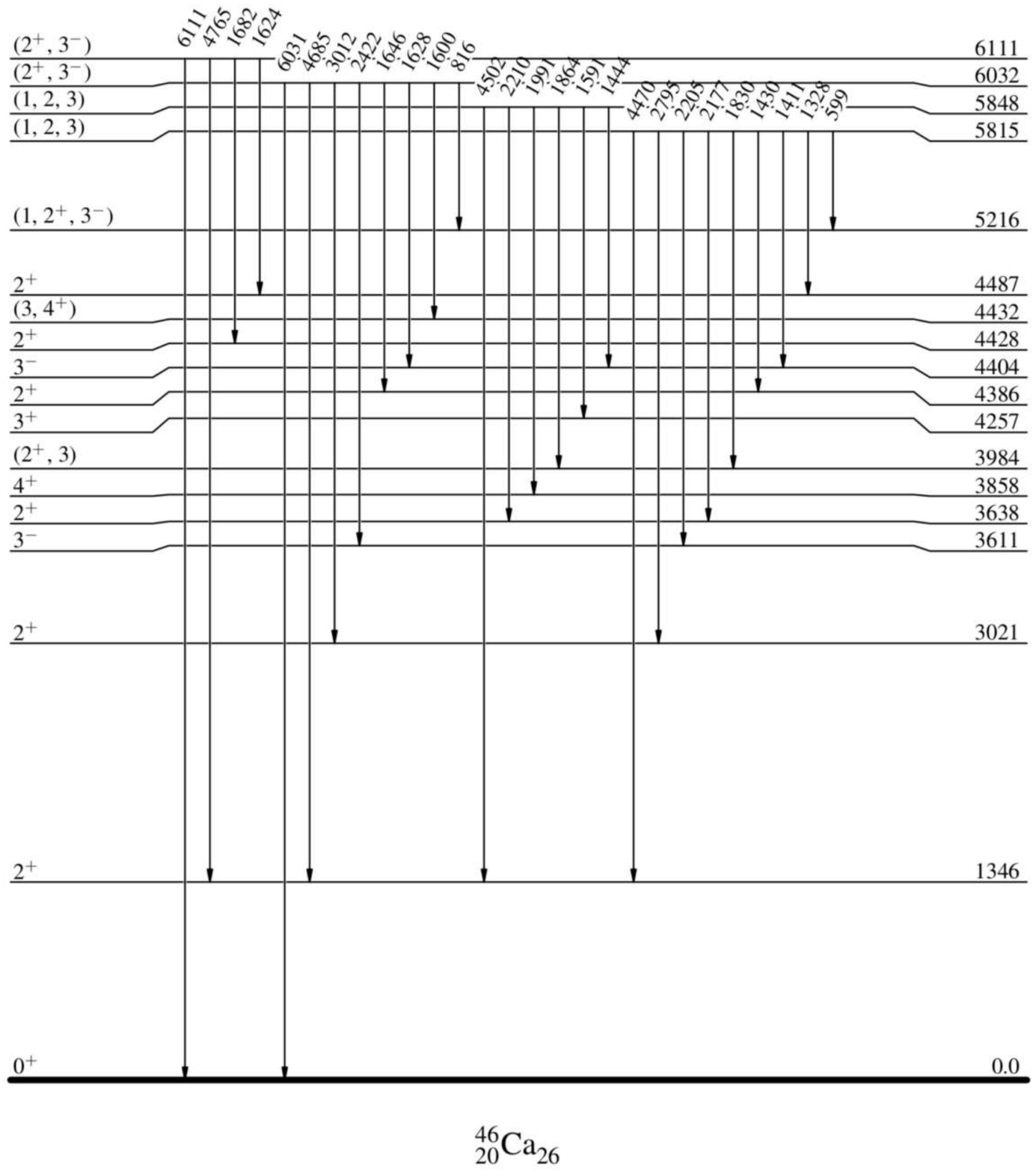}
		\caption{A partial level scheme of levels populated in $^{46}$Ca from the $\beta^-$ decay of the $J^{\pi}$ = 2$^- $ $^{46}$K ground state showing $\gamma$ rays that depopulate the 5815-6111 keV levels.}
	
\end{figure*}

\begin{figure*}
\centering
    
		\includegraphics[scale=0.8]{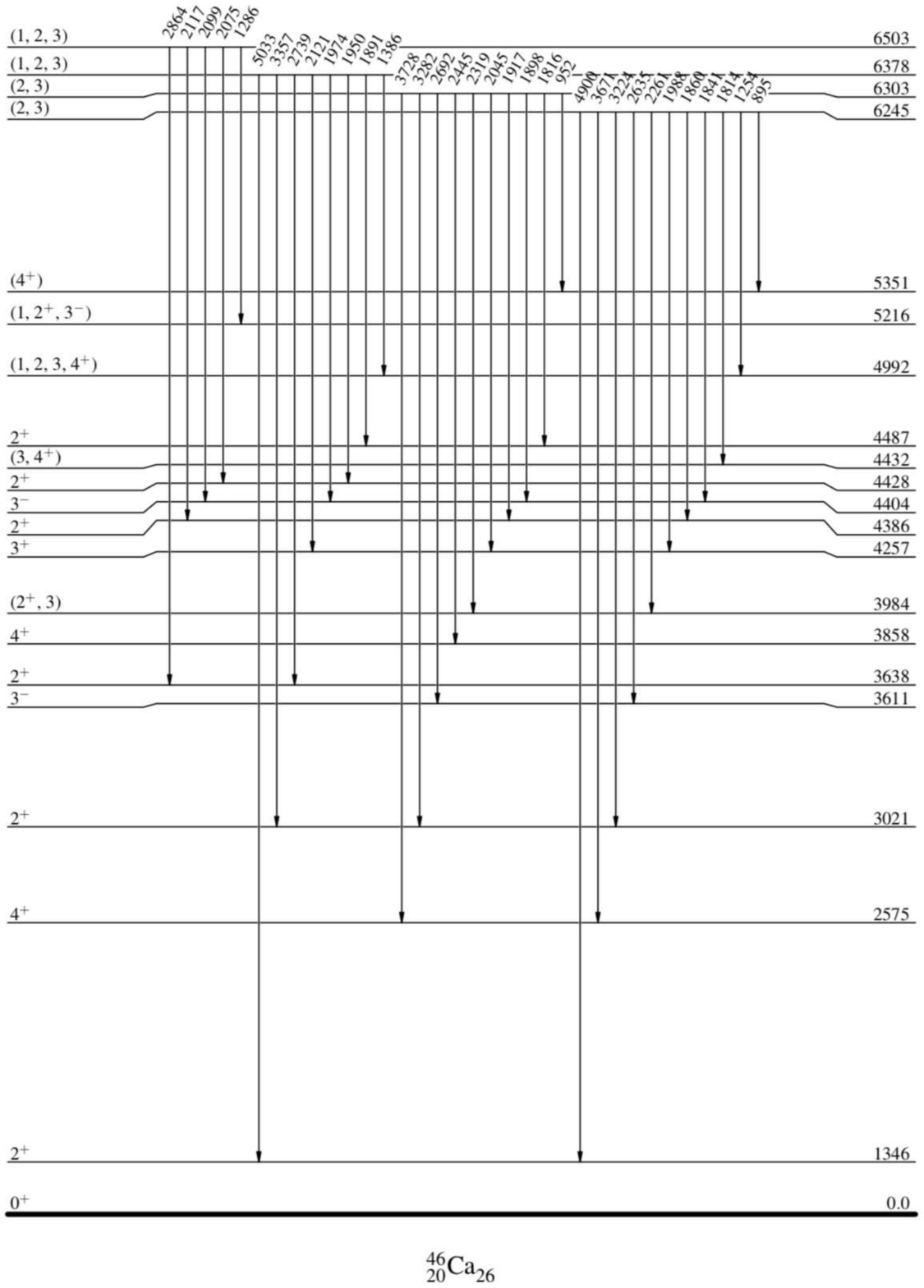}
		\caption{A partial level scheme of levels populated in $^{46}$Ca from the $\beta^-$ decay of the $J^{\pi}$ = 2$^- $ $^{46}$K ground state showing $\gamma$ rays that depopulate the 6245-6503 keV levels.}
	
\end{figure*}

\begin{figure*}
\centering
    
		\includegraphics[scale=0.8]{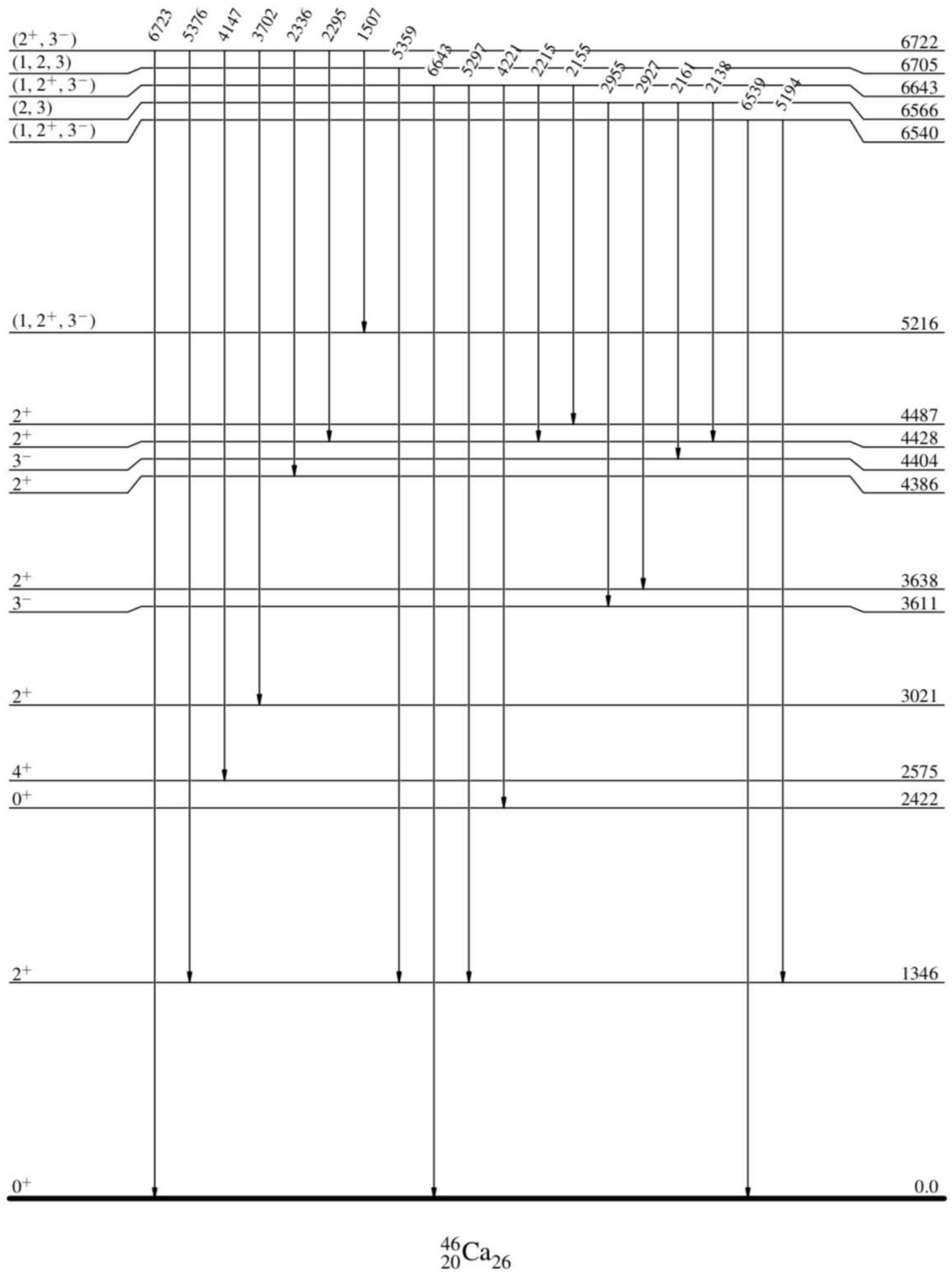}
		\caption{A partial level scheme of levels populated in $^{46}$Ca from the $\beta^-$ decay of the $J^{\pi}$ = 2$^- $ $^{46}$K ground state showing $\gamma$ rays that depopulate the 6540-6723 keV levels.}
	
\end{figure*}

\end{document}